\pdfoutput=1
\documentclass{article}
\usepackage{iclr2027_conference,times}
\iclrfinalcopy

\usepackage{amsmath,amsfonts,bm}

\def\eqref#1{equation~\ref{#1}}

\def\1{\bm{1}}

\DeclareMathAlphabet{\mathsfit}{\encodingdefault}{\sfdefault}{m}{sl}
\SetMathAlphabet{\mathsfit}{bold}{\encodingdefault}{\sfdefault}{bx}{n}

\usepackage{doi}
\usepackage{hyperref,url,graphicx,booktabs,amsmath,amssymb,makecell,adjustbox,subcaption,xcolor}
\usepackage{multirow,array,float,caption,wrapfig,needspace,colortbl}
\usepackage{placeins}


\makeatletter
\newcommand{\wrapfill}{\par\ifnum\value{WF@wrappedlines}>0 \addtocounter{WF@wrappedlines}{-1}\vspace{\value{WF@wrappedlines}\baselineskip}\setcounter{WF@wrappedlines}{0}\WFclear\fi}
\makeatother
\definecolor{bestc}{HTML}{CFE2F3}\definecolor{secondc}{HTML}{D9EAD3}
\newcommand{\best}[1]{\cellcolor{bestc}#1}\newcommand{\second}[1]{\cellcolor{secondc}#1}
\newcommand{\method}{JamTet}

\title{Physics-Based Sphere Packing for\\ Lagrangian Mesh Morphing}
\author{Jiong Lin \& Hod Lipson\\Columbia University}
\begin{document}
\maketitle
\lhead{Preprint.}
\vspace{-6pt}
\begin{figure}[H]
  \centering
\setlength{\tabcolsep}{0pt}
\begin{tabular}{@{}c@{}}
\includegraphics[width=0.1650\textwidth]{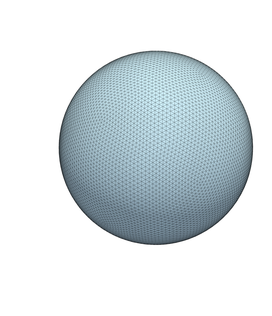}\includegraphics[width=0.1650\textwidth]{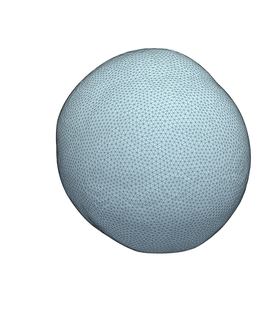}\includegraphics[width=0.1650\textwidth]{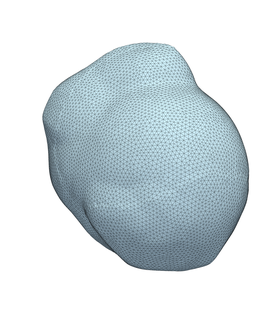}\includegraphics[width=0.1650\textwidth]{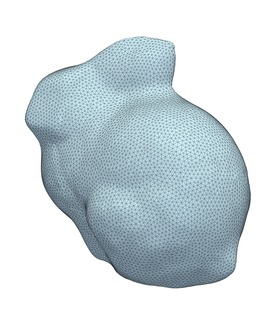}\includegraphics[width=0.1650\textwidth]{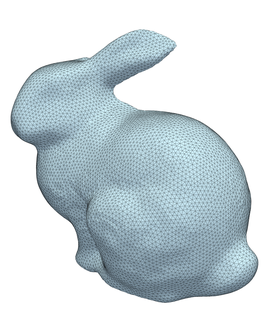}\includegraphics[width=0.1650\textwidth]{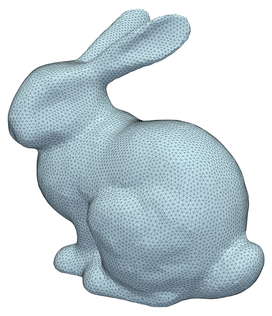} \\[2pt]
\includegraphics[width=0.1650\textwidth]{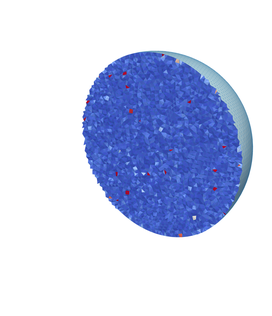}\includegraphics[width=0.1650\textwidth]{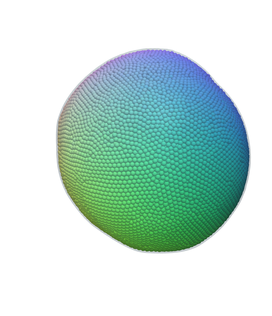}\includegraphics[width=0.1650\textwidth]{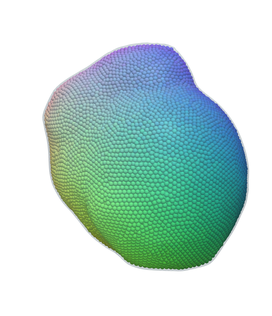}\includegraphics[width=0.1650\textwidth]{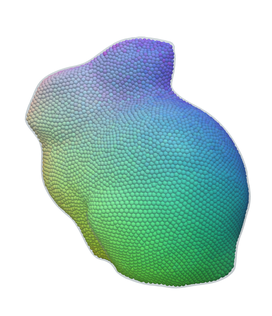}\includegraphics[width=0.1650\textwidth]{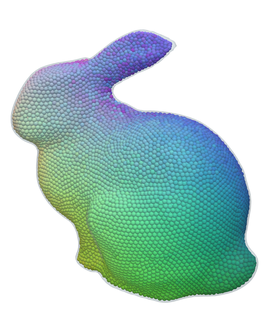}\includegraphics[width=0.1650\textwidth]{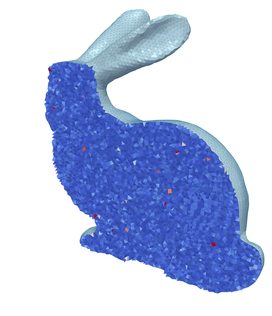} \\
\end{tabular}

  \caption{Sphere-to-bunny morph that keeps sphere identities throughout. Top: surface mesh per step; bottom: the spheres, coloured by starting position, settled in each surface and connected into tetrahedra; initial and target meshes cut open, coloured by element quality: regular \raisebox{-0.15ex}{\includegraphics[width=3em,height=1.1ex]{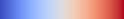}} distorted.}
  \label{fig:teaser}
\end{figure}
\vspace{-4pt}
\begin{abstract}
This paper studies tetrahedral meshes as the body representation for differentiable simulation and computational design. Fixed-connectivity meshes degrade under large morphs, while remeshing from scratch discards node correspondence. We present \method{}, a physics-based sphere-packing framework for volumetric meshing and morphing. We contribute (i) a GPU-parallel mesher combining octree-hierarchical packing with constrained Delaunay tetrahedralization, producing more uniform element volumes than TetGen and fTetWild; (ii) Lagrangian mesh morphing that preserves interior-node identities by re-equilibrating the same spheres within changing shapes and rebuilding the boundary and connectivity, remaining inversion-free where fixed-connectivity and TetSphere meshes invert; and (iii) a differentiable GPU simulator in JAX, with mass--spring edges and a volumetric Neo-Hookean term, integrated with mesh morphing in a design pipeline. In soft-robot morphology design experiments, interior-node gradients improve swimming fitness by $0.73$--$1.07$ over a matched surface-only variant, while voxelized versions of the same designs yield 32--63\% lower fitness. These results establish sphere packing as a practical volumetric mesh representation for gradient-based shape optimization. Code and media: under review.
\end{abstract}

\section{Introduction}
\label{sec:intro}
Gradient-based shape optimization with differentiable physics changes the geometry of a body step by step, and the volumetric mesh used by the simulator has to conform to each new boundary with well-shaped tetrahedra. If its interior nodes also keep their identities, values stored on those nodes carry over from one shape to the next. TetSphere splatting~\citep{guo2024tetsphere} keeps node identity with a Lagrangian volumetric mesh, a union of tetrahedralized spheres whose nodes move with the shape, and discourages inverted elements with a soft penalty; in our morph experiments, some of its tetrahedra still invert. Tetrahedral meshers such as TetGen~\citep{si2015tetgen} and fTetWild~\citep{hu2020ftetwild} conform to a given boundary and control element quality, and each call builds a new mesh with new node indices. A mesh that changes shape thus has two standard options: fixed connectivity keeps node identity while its elements distort, inverting 4.1\% of the tetrahedra by the end of a sphere-to-bunny morph, and remeshing from scratch restores element quality while assigning new node indices (Section~\ref{sec:exp:morph}). We target a conforming tetrahedral mesh whose interior nodes persist and whose elements stay well shaped through large shape changes.

Soft-robot morphology design is one application that needs such meshes. Evolutionary design grows soft bodies from voxels~\citep{cheney2013unshackling,bhatia2021evogym}, and recent systems design voxel robots with learned controllers~\citep{li2025endoskeletal} or pretrain them in differentiable simulation~\citep{strgar2026morphological}. ChainQueen~\citep{hu2019chainqueen} and DiffTaichi~\citep{hu2020difftaichi} simulate soft robots with differentiable particle and mass--spring models. Voxel bodies keep one connectivity and a boundary of cube faces, and particle bodies have no element connectivity (Appendix~\ref{app:repr}).

\begin{figure}[t]
  \centering
  \setlength{\tabcolsep}{0pt}%
  \begin{tabular}{@{}l@{\hspace{3pt}}l@{}}
  {\small (a)} & \raisebox{-\height}{\includegraphics[width=\dimexpr\textwidth-16pt\relax]{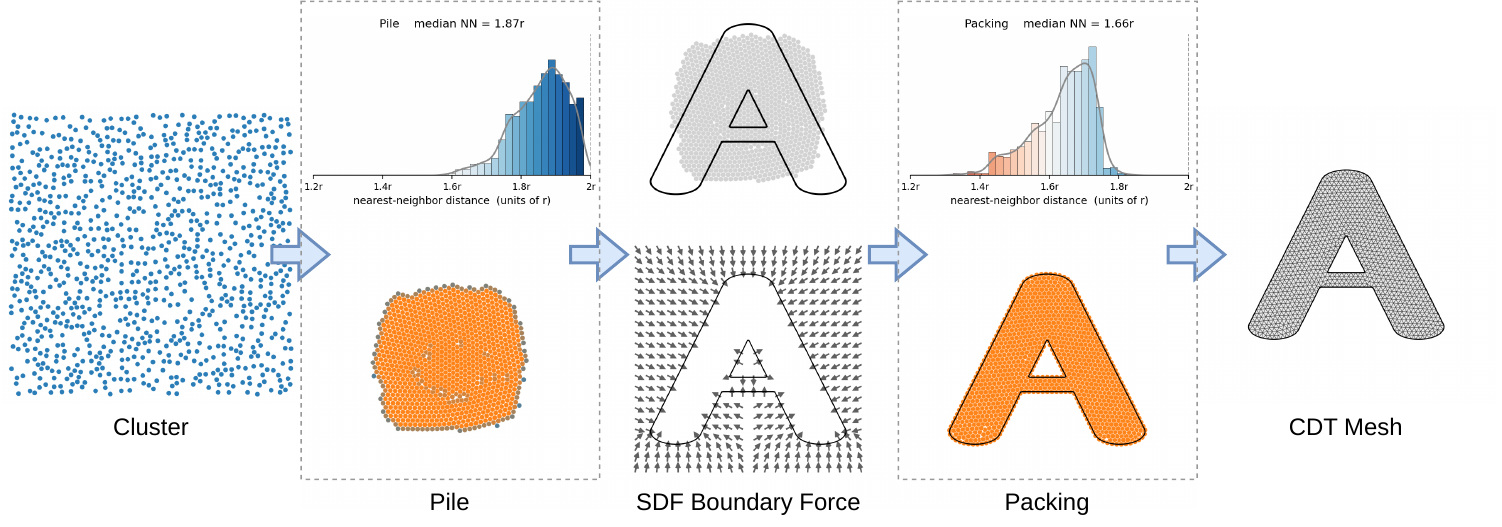}}\\[4pt]
  {\small (b)} & \raisebox{-\height}{\includegraphics[width=\dimexpr\textwidth-16pt\relax]{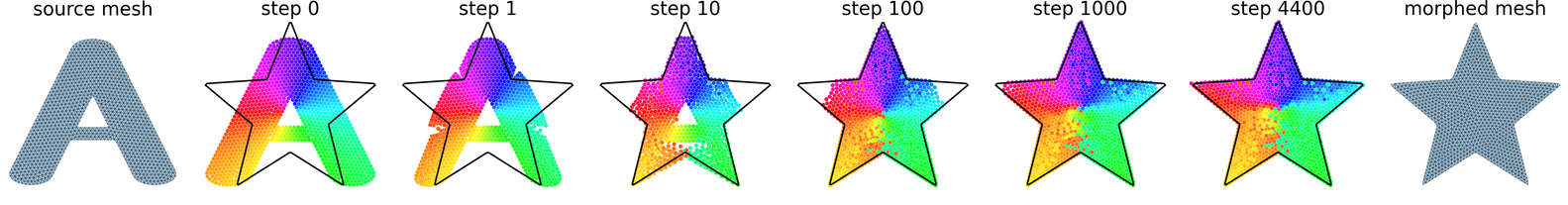}}
  \end{tabular}
  \caption{Sphere packing mesh in 2D. (a)~Mesh generation, left to right: a loose cluster settles into a pile, a boundary force pushes it into the target shape, relaxation gives a tight packing (insets: neighbour distances), and constrained Delaunay triangulation gives the mesh. (b)~Morphing: a packed letter ``A'' morphs into a star, and the balls, coloured by starting position (hue: angle; brightness: radius), keep their identities.}
  \label{fig:overview2d}
\end{figure}

\method{} builds the mesh from physical spheres (Figure~\ref{fig:overview2d}a). Given a surface, spheres flow inside it like a liquid and then jam into a dense packing~\citep{ohern2003jamming}; their centres become the interior nodes, and constrained Delaunay tetrahedralization (CDT) connects them with the nodes of the re-triangulated surface. To morph, the same spheres re-settle inside each new surface and CDT rebuilds the connectivity, so every interior node keeps its identity (Figure~\ref{fig:overview2d}b). A differentiable GPU simulator in JAX runs on these meshes, and a design loop turns its gradients into new shapes through the morph step, so every design keeps the same interior spheres.

We make three contributions.
(i)~A sphere-packing mesher with GPU packing, an octree schedule and CDT connectivity, whose meshes match or exceed the element quality of TetGen and fTetWild with more uniform element volumes (Section~\ref{sec:exp:static}).
(ii)~Identity-preserving morphing: along a sphere-to-bunny morph, the mesh keeps every interior node identity and has no inverted tetrahedron at any step (Section~\ref{sec:exp:morph}).
(iii)~A differentiable GPU simulator with mass--spring edges and a volumetric Neo-Hookean term, coupled with the morph step in a morphology-design pipeline. For soft swimmers, gradients on interior nodes raise fitness by 0.73--1.07 over a matched surface-only variant in all four swimming conditions, and voxel versions of the designs reach 32--63\% lower fitness than the smooth designs (Section~\ref{sec:exp:robot}).

Section~\ref{sec:related} reviews related work, Section~\ref{sec:method} describes the method, Section~\ref{sec:exp} reports the experiments, and Section~\ref{sec:conclusion} discusses limitations and concludes.

\section{Related Work}
\label{sec:related}
\textbf{Tetrahedral meshing.} Delaunay refinement~\citep{shewchuk1998refinement} and constrained Delaunay tetrahedralization, as in TetGen~\citep{si2015tetgen}, recover a given boundary and insert Steiner points to bound element quality. fTetWild~\citep{hu2020ftetwild} meshes imperfect input within a tolerance envelope and then optimizes element quality. Among particle methods, bubble packing relaxes bubbles with simulated forces before Delaunay triangulation~\citep{shimada1995bubble}, ellipsoidal bubble packing connects the nodes by an advancing front for anisotropic meshes~\citep{yamakawa2000ellipsoidal}, later sphere packings vary the element size or pack faster~\citep{loWang2005advancing,liuLi2008fourR,qi2014accelerated}, and DistMesh balances spring forces inside a signed distance field~\citep{persson2004distmesh}. \method{} builds on sphere packing with a two-stage jamming schedule, an octree and GPU execution, and it carries one packing across shape changes.

\textbf{Deforming volumetric representations.} The material point method (MPM)~\citep{sulsky1994mpm,hu2018mlsmpm} carries a deforming body on particles and solves its motion on a background grid, and ChainQueen~\citep{hu2019chainqueen} differentiates it for soft-robot design. A tetrahedral mesh gives the body an explicit boundary and elements on which finite-element analysis runs directly; Appendix~\ref{app:fea} uses our meshes for compliance design with linear finite elements. DefTet~\citep{gao2020deftet} deforms a fixed tetrahedral grid and predicts per-cell occupancy. TetSphere splatting~\citep{guo2024tetsphere} represents a shape as a union of deformable tetrahedralized spheres whose nodes persist during optimization, and it is the closest prior representation. DMTet~\citep{shen2021dmtet}, FlexiCubes~\citep{shen2023flexicubes} and TetWeave~\citep{binninger2025tetweave} extract surfaces from fields stored on deformable tetrahedral, cubical or Delaunay background grids, which TetWeave rebuilds from optimized points every few steps. Level-set methods morph solids by evolving an implicit function~\citep{breen2001levelset}, and our morphs follow such a path.

\textbf{Differentiable simulation and morphology design.} ChainQueen~\citep{hu2019chainqueen}, DiffTaichi~\citep{hu2020difftaichi}, PlasticineLab~\citep{huang2021plasticinelab} and DiffPD~\citep{du2021diffpd} differentiate soft-body rollouts with respect to states and parameters. DiffAqua~\citep{ma2021diffaqua} optimizes soft swimmer shapes with such gradients. ELDiR~\citep{strgar2024evolution} evolves robot bodies while training their neural controllers through a differentiable mass--spring simulator on the GPU. Appendix~\ref{app:repr} compares the body representations of these systems.

\section{Method}
\label{sec:method}
Section~\ref{sec:method:physics} packs spheres inside a surface and connects them into tetrahedra, Section~\ref{sec:method:schedule} accelerates the packing, Section~\ref{sec:method:morph} carries the spheres through a sequence of shapes, and Section~\ref{sec:method:sim} simulates the meshes and turns gradients into shape updates. Throughout, $\Omega$ is the domain inside a closed surface with signed distance function $\phi$ ($\phi<0$ inside), $n$ the number of interior spheres, $s$ the node spacing, $\mathcal{S}$ the surface nodes, and $\mathbf{p}_i$ and $\mathbf{v}_i$ the centre and velocity of sphere $i$, with $\mathcal{P}=\{\mathbf{p}_i\}_{i=1}^{n}$.

\subsection{Surface-first sphere packing}
\label{sec:method:physics}
One length sets the resolution: the node spacing $s = 0.93\,(|\Omega|/n)^{1/3}$, which matches the median nearest-neighbour distance of our packings. Every sphere has radius $r = (3|\Omega|/4\pi n)^{1/3} \approx 0.667\,s$, so the spheres together have the volume of $\Omega$ and overlap their neighbours. Meshing runs in three steps (Figure~\ref{fig:overview2d}a). First, the boundary is re-triangulated with isotropic triangles of edge length $s$~\citep{botsch2004remeshing}, whose nodes $\mathcal{S}$ stay fixed, and $\phi$ is recomputed from this surface on a $128^3$ grid. Second, $n$ spheres are packed inside it. Third, CDT connects $\mathcal{S}\cup\mathcal{P}$ into tetrahedra.

Packing moves the centres under two pair forces applied in sequence. The \emph{water} stage uses a long-range Gaussian repulsion,
\begin{equation}
\mathbf{f}^{\mathrm{w}}_i = k_{\mathrm{w}} \sum_{j\neq i} \exp\!\big(-\|\mathbf{p}_i-\mathbf{p}_j\|^2/s^2\big)\,\hat{\mathbf{u}}_{ij},
\label{eq:water}
\end{equation}
where $\hat{\mathbf{u}}_{ij}$ is the unit vector from $\mathbf{p}_j$ to $\mathbf{p}_i$. The force acts at every distance up to its cutoff $2.2\,s$, so spheres keep sliding past one another and flow into empty regions. The stage runs only while the packing has a hole, that is, while some interior point deeper than $s$ lies farther than $1.6\,s$ from every centre. The \emph{jam} stage uses a repulsion-only contact force,
\begin{equation}
\mathbf{f}^{\mathrm{j}}_i = k_{\mathrm{c}} \sum_{j\neq i} \max\big(2r-\|\mathbf{p}_i-\mathbf{p}_j\|,\,0\big)\,\hat{\mathbf{u}}_{ij},
\label{eq:jam}
\end{equation}
which acts only between overlapping spheres. Each sphere becomes caged by its neighbours, and the damped motion settles into a jammed packing; the stage stops when the mean speed falls below $10^{-4}$. Both stages integrate
\begin{equation}
\mathbf{v}_i \leftarrow \gamma\,\mathbf{v}_i + \Delta t\,\mathbf{f}_i, \qquad
\mathbf{p}_i \leftarrow \Pi\big(\mathbf{p}_i + \Delta t\,\mathbf{v}_i\big),
\label{eq:update}
\end{equation}
with unit mass, $\gamma = 0.82$, $\Delta t = 0.005$, $k_{\mathrm{w}} = 300$ and $k_{\mathrm{c}} = 500$, and each step is capped at $\|\mathbf{v}_i\|\,\Delta t \le c\,s$, with $c=0.5$ for jam and $c=0.12$ for water.

The projection $\Pi$ keeps every centre at least $s$ inside the surface. It moves each centre with $\phi(\mathbf{p}) > -s$ onto the level set $\phi = -s$ by five Newton steps $\mathbf{p} \leftarrow \mathbf{p} - (\phi(\mathbf{p}) + s)\,\nabla\phi / \|\nabla\phi\|$ and removes the outward normal component of its velocity. The first interior layer therefore sits one spacing below the surface, and boundary tetrahedra have the size of interior ones. At full resolution, the surface nodes $\mathcal{S}$ enter Eq.~(\ref{eq:jam}) as fixed spheres, so the first layer settles into the hollows above the surface triangles. TetGen~\citep{si2015tetgen} then computes the CDT of $\mathcal{S}\cup\mathcal{P}$ with the surface triangles as constraints, and the vertices of the mesh are exactly these nodes. The mesh stores the surface nodes first and the interior spheres next, in a fixed order, so the $i$-th interior node is sphere $i$. The design loop adds TetGen's quality refinement (radius--edge bound 1.414), which appends helper nodes after the input nodes, so the interior block is unchanged.

\subsection{Octree-hierarchical packing}
\label{sec:method:schedule}
Direct packing of $n=64$k spheres runs for thousands of steps, and each step rebuilds a spatial hash to find neighbours. The octree schedule builds the packing from coarse to fine (Figure~\ref{fig:octree}). Level~0 packs at most $n/64$ spheres at spacing $4s$ with the two stages of Section~\ref{sec:method:physics}. Each finer level $\ell$, with spacing $s_\ell$, replaces every sphere by eight children at the corners of a cube of side $s_\ell$ centred on it. It then injects spheres at grid points farther than $0.95\,s_\ell$ from every centre, up to the level's target count, which fills thin features that no coarse sphere reached. The children start near equilibrium and move less than one spacing, so each level builds one neighbour list, refreshes it every 200 steps, and runs the jam stage on this cached list; the water stage runs only if the fill check fails. A step then costs one gather over a fixed index array.

\subsection{Morphing with persistent spheres}
\label{sec:method:morph}
A morph moves the same spheres through a sequence of shapes. A level-set flow~\citep{breen2001levelset} carries the source $\phi_0$ towards the target $\phi_1$,
\begin{equation}
\partial_\tau \phi = \phi_1\,\|\nabla\phi\|,
\label{eq:levelset}
\end{equation}
which grows the shape where it lies inside the target and shrinks it outside, so the surface approaches the target monotonically. We take snapshots evenly spaced in the resolved fraction of the source--target symmetric-difference volume. Each snapshot surface is extracted, smoothed~\citep{taubin1995signal}, scaled to the target volume so that the fixed set of spheres keeps spacing $s$, and re-triangulated. The spheres of the previous step are projected inside the new surface, the water stage carries them into new volume until the fill check passes, and the jam stage settles them against the new surface nodes. CDT then rebuilds the connectivity. The number of spheres stays fixed, so interior node $i$ at every step is sphere $i$ of the source packing, while the surface is re-triangulated at every step.

\newcommand{\OctreeBlock}{%
\begin{wrapfigure}{r}{\dimexpr0.5\textwidth-\columnsep\relax}
  \centering
\providecommand{\octlab}[1]{{\scriptsize #1}}
\begingroup\setlength{\tabcolsep}{0pt}\renewcommand{\arraystretch}{0}%
\begin{tabular}{@{}ccc@{}}
\includegraphics[width=0.333\linewidth]{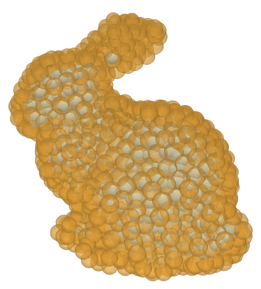} &
\includegraphics[width=0.333\linewidth]{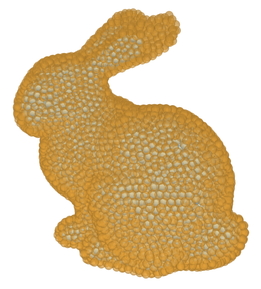} &
\includegraphics[width=0.333\linewidth]{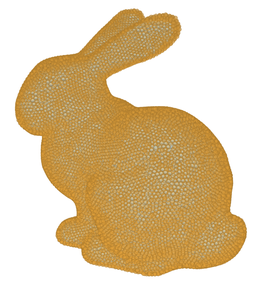} \\
\includegraphics[width=0.333\linewidth]{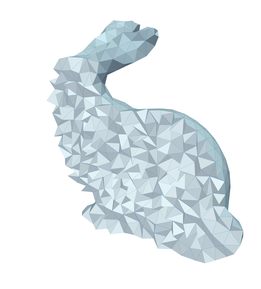} &
\includegraphics[width=0.333\linewidth]{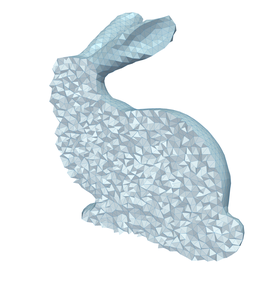} &
\includegraphics[width=0.333\linewidth]{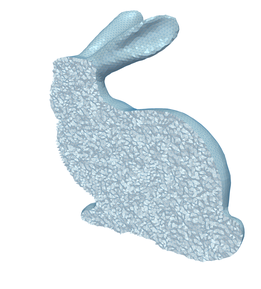} \\[2pt]
\renewcommand{\arraystretch}{1}\octlab{level 0} & \octlab{level 1} & \octlab{level 2} \\
\end{tabular}\par\vspace{3pt}
\includegraphics[width=0.49\linewidth]{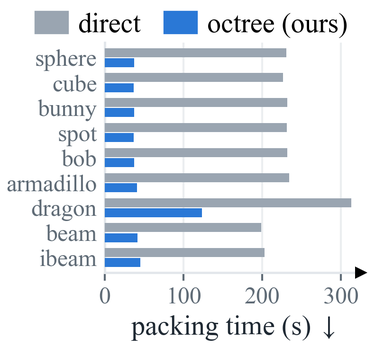}\hfill
\includegraphics[width=0.49\linewidth]{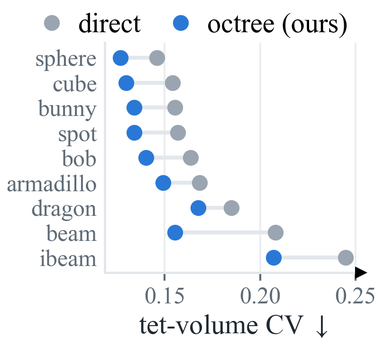}
\endgroup

  \caption{Octree packing on the bunny. Top: three levels (each ball split into 8), packing and cut mesh; bottom: packing time and coefficient of variation (CV) of tetrahedron volumes at 64k nodes per shape, octree vs.\ direct.}
  \label{fig:octree}
  \vspace{8pt}
  {\captionof{table}{Static mesh quality at 64k interior spheres, per shape; measures defined in Section~\ref{sec:exp:static}; no inverted elements for any method; all nine shapes: Table~\ref{tab:app-static}.\protect\footnotemark}
  \label{tab:static-compare-64k}
  \resizebox{\linewidth}{!}{
\begin{tabular}{@{}l c c c c r@{}}
\toprule
Method & Vol\,CV & AMIPS & Min Dih & Aspect & Wall \\
 & {\scriptsize $(\sigma/\mu)\downarrow$} & {\scriptsize Avg$\pm$std$\downarrow$} & {\scriptsize deg$\uparrow$} & {\scriptsize Avg$\pm$std$\uparrow$} & {\scriptsize (s)} \\
\midrule
\multicolumn{6}{@{}l}{\emph{Bunny}} \\
TetGen & \second{$0.296$} & \second{$3.81 \pm 0.88$} & $48.6 \pm 12.6$ & $0.573 \pm 0.133$ & \best{$0.5$} \\
fTetWild & $0.540$ & \best{$3.33 \pm 0.26$} & \second{$56.0 \pm 7.3$} & \second{$0.718 \pm 0.137$} & $77.2$ \\
Ours & \best{$0.134$} & \best{$3.33 \pm 0.90$} & \best{$57.0 \pm 9.2$} & \best{$0.733 \pm 0.128$} & \second{$36.9$} \\
\midrule
\multicolumn{6}{@{}l}{\emph{Spot}} \\
TetGen & \second{$0.296$} & $3.80 \pm 0.88$ & $48.7 \pm 12.6$ & $0.574 \pm 0.133$ & \best{$0.5$} \\
fTetWild & $0.332$ & \second{$3.49 \pm 0.30$} & \second{$53.1 \pm 8.6$} & \second{$0.621 \pm 0.112$} & $50.3$ \\
Ours & \best{$0.134$} & \best{$3.33 \pm 0.87$} & \best{$56.9 \pm 9.2$} & \best{$0.732 \pm 0.127$} & \second{$36.4$} \\
\bottomrule
\end{tabular}
}\par}
\end{wrapfigure}
}
\subsection{Differentiable simulation and design}
\label{sec:method:sim}
The mesh carries a mass--spring model with a volumetric Neo-Hookean term. Every tetrahedron edge $e=(i,j)$ is a damped spring with rest length $L_e$, and every tetrahedron $\tau$ with rest volume $V_\tau$ adds a Neo-Hookean volume energy in its volume ratio $J_\tau$:
\begin{equation}
E(\mathbf{x}) = \sum_{e} \tfrac{1}{2}\,k_s\big(\|\mathbf{x}_i-\mathbf{x}_j\| - a_e(t)\,L_e\big)^2 + \sum_{\tau} \mu\,V_\tau \big(J_\tau - 1 - \ln J_\tau\big).
\label{eq:energy}
\end{equation}

\begin{wrapfigure}{r}{\dimexpr0.5\textwidth-\columnsep\relax}
  \centering
\providecommand{\octlab}[1]{{\scriptsize #1}}
\begingroup\setlength{\tabcolsep}{0pt}\renewcommand{\arraystretch}{0}%
\begin{tabular}{@{}ccc@{}}
\includegraphics[width=0.333\linewidth]{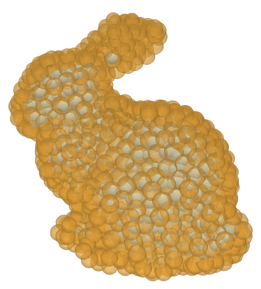} &
\includegraphics[width=0.333\linewidth]{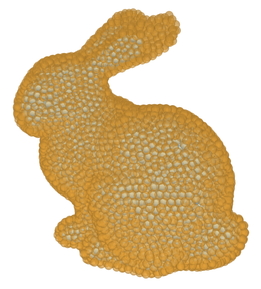} &
\includegraphics[width=0.333\linewidth]{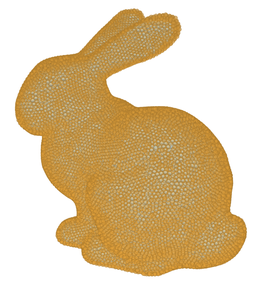} \\
\includegraphics[width=0.333\linewidth]{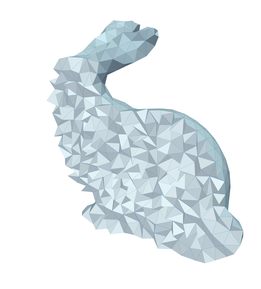} &
\includegraphics[width=0.333\linewidth]{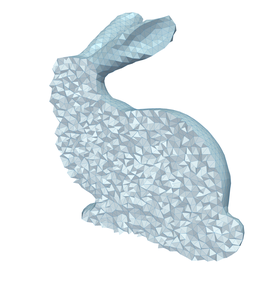} &
\includegraphics[width=0.333\linewidth]{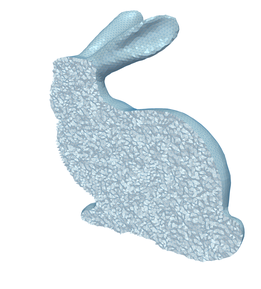} \\[2pt]
\renewcommand{\arraystretch}{1}\octlab{level 0} & \octlab{level 1} & \octlab{level 2} \\
\end{tabular}\par\vspace{3pt}
\includegraphics[width=0.49\linewidth]{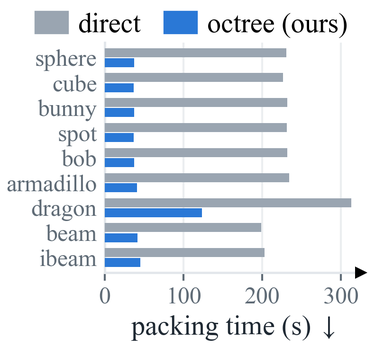}\hfill
\includegraphics[width=0.49\linewidth]{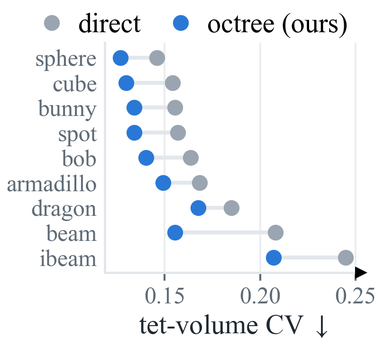}
\endgroup

  \caption{Octree packing on the bunny. Top: three levels (each ball split into 8), packing and cut mesh; bottom: packing time and coefficient of variation (CV) of tetrahedron volumes at 64k nodes per shape, octree vs.\ direct.}
  \label{fig:octree}
  \vspace{8pt}
  {\captionof{table}{Static mesh quality at 64k interior spheres, per shape; measures defined in Section~\ref{sec:exp:static}; no inverted elements for any method; all nine shapes: Table~\ref{tab:app-static}.\protect\footnotemark}
  \label{tab:static-compare-64k}
  \resizebox{\linewidth}{!}{
\begin{tabular}{@{}l c c c c r@{}}
\toprule
Method & Vol\,CV & AMIPS & Min Dih & Aspect & Wall \\
 & {\scriptsize $(\sigma/\mu)\downarrow$} & {\scriptsize Avg$\pm$std$\downarrow$} & {\scriptsize deg$\uparrow$} & {\scriptsize Avg$\pm$std$\uparrow$} & {\scriptsize (s)} \\
\midrule
\multicolumn{6}{@{}l}{\emph{Bunny}} \\
TetGen & \second{$0.296$} & \second{$3.81 \pm 0.88$} & $48.6 \pm 12.6$ & $0.573 \pm 0.133$ & \best{$0.5$} \\
fTetWild & $0.540$ & \best{$3.33 \pm 0.26$} & \second{$56.0 \pm 7.3$} & \second{$0.718 \pm 0.137$} & $77.2$ \\
Ours & \best{$0.134$} & \best{$3.33 \pm 0.90$} & \best{$57.0 \pm 9.2$} & \best{$0.733 \pm 0.128$} & \second{$36.9$} \\
\midrule
\multicolumn{6}{@{}l}{\emph{Spot}} \\
TetGen & \second{$0.296$} & $3.80 \pm 0.88$ & $48.7 \pm 12.6$ & $0.574 \pm 0.133$ & \best{$0.5$} \\
fTetWild & $0.332$ & \second{$3.49 \pm 0.30$} & \second{$53.1 \pm 8.6$} & \second{$0.621 \pm 0.112$} & $50.3$ \\
Ours & \best{$0.134$} & \best{$3.33 \pm 0.87$} & \best{$56.9 \pm 9.2$} & \best{$0.732 \pm 0.127$} & \second{$36.4$} \\
\bottomrule
\end{tabular}
}\par}
\end{wrapfigure}
The tetrahedral mesh defines this model directly: its nodes carry the mass, its edges form the spring network, and the jammed packing gives edges of similar length. A spring network has no measure of volume, so under large deformation a tetrahedron can collapse or invert. The well-shaped tetrahedra of our meshes therefore also carry the Neo-Hookean volume term, whose energy rises steeply as $J_\tau$ approaches zero; it resists collapse and inversion and approximates the near incompressibility of soft tissue. The activation $a_e(t)$ scales the rest length of muscle edges and equals one for passive edges. In the implementation, $J_\tau$ is clamped at 0.01, and $\mu$ is scaled down for tetrahedra whose rest volume lies below 10\% of the median. Semi-implicit Euler integrates the nodes, which carry equal masses. For swimming, water acts through surface forces~\citep{min2019softcon,ma2021diffaqua}. We apply a linear drag $-A\,(c_n\mathbf{v}_n + c_t\mathbf{v}_t)$ to each boundary triangle with area $A$, where $\mathbf{v}_n$ and $\mathbf{v}_t$ are the normal and tangential parts of its velocity and $c_n = 6\,c_t$. The whole simulator is written in JAX~\citep{jax2018github} and runs on the GPU as one compiled scan over time steps, fast enough to iterate on designs: a 6\,s rollout takes 0.4--2.3\,s from 1k to 64k spheres (Appendix~\ref{app:simspeed}). Automatic differentiation gives the gradient of a rollout objective with respect to every node's rest position in 3.6--6.3\,s at 1k and 8k, so the morphology optimization uses simulator gradients directly. The backward pass recomputes each block of steps from a checkpoint (500 steps at 1k spheres, 125 at 8k), so memory grows with the number of blocks, and at each step it clips the norm of the state cotangent to $10^3$ for numerical stability.

\textbf{Design loop.} Each design iteration turns the fitness of a rollout into a new shape in four steps. (1)~The simulator computes the gradient of the fitness with respect to the rest position of every node, averaged over eight perturbed rollouts. (2)~The node gradients are smoothed on a grid and transferred to the design parameters, the displacement vectors of a free-form deformation (FFD) lattice~\citep{sederberg1986ffd} that warps the level set of the body. (3)~A heavy-ball step with momentum 0.9 updates these displacements. (4)~The morph step of Section~\ref{sec:method:morph} re-packs the spheres inside the warped shape and rebuilds the tetrahedral mesh with hybrid CDT. Gradients reach the node positions through the simulator, while the morph step maps each level set to a valid mesh and passes no gradient.\footnotetext{Shading in all tables: \colorbox{bestc}{best}, \colorbox{secondc}{second best}.}

\section{Experiments}
\label{sec:exp}
We evaluate \method{} in three settings: static meshing of nine shapes against TetGen and fTetWild (Section~\ref{sec:exp:static}), morphing against fixed connectivity, fTetWild and TetSphere (Section~\ref{sec:exp:morph}), and soft-robot morphology design with the differentiable simulator (Section~\ref{sec:exp:robot}). The appendix gives the full tables and settings (Appendices~\ref{app:static}--\ref{app:voxel}) and a finite-element design study (Appendix~\ref{app:fea}).

\subsection{Static meshing}
\label{sec:exp:static}
We mesh nine shapes, from a sphere, a cube and two beams to the bunny, Spot, Bob, Armadillo and the dragon, with $n = 64$k interior spheres. TetGen (switches \texttt{pYq1.414a}) and fTetWild mesh the same shapes, with the maximum tetrahedron volume or the edge length tuned to our node count. Our packing runs on one GPU; our surface triangulation and CDT, TetGen and fTetWild run on the CPU. Table~\ref{tab:static-compare-64k} reports five measures: \emph{Vol\,CV}, the coefficient of variation of tetrahedron volumes; \emph{AMIPS}, the mean AMIPS energy~\citep{fu2015amips}, 3 for a regular tetrahedron; \emph{Min Dih}, the smallest dihedral angle of each tetrahedron averaged over the mesh, 70.5$^\circ$ when regular; \emph{Aspect}, a normalized aspect ratio, 1 when regular; and \emph{Wall}, the time of the whole meshing pipeline.

On the bunny and Spot, our meshes rank first in all four quality measures, with AMIPS tied with fTetWild on the bunny, and second in wall time (Table~\ref{tab:static-compare-64k}). Their Vol\,CV is 45\% of that of TetGen and 25--40\% of that of fTetWild. No mesher produces an inverted element. Over all nine shapes, our meshes have the lowest Vol\,CV on every shape (Appendix~\ref{app:static}).

The low volume spread is consistent with the packing, whose jam stage leaves neighbouring centres at nearly equal distances. The octree schedule (Figure~\ref{fig:octree}) packs 2.5--6.3$\times$ faster than direct packing on the nine shapes and gives a lower Vol\,CV on every shape.

\begin{figure}[t]
  \centering
  \begin{minipage}[c]{0.60\textwidth}
\providecommand{\morphlab}[1]{{\scriptsize #1}}%
\ifdefined\morphtile\else\newlength{\morphtile}\fi%
\setlength{\tabcolsep}{1pt}\renewcommand{\arraystretch}{0.75}%
\setlength{\morphtile}{\dimexpr(\linewidth-0.17\linewidth-12\tabcolsep)/6\relax}%
\begin{tabular}{@{}c@{}}
\begin{tabular}[c]{@{}>{\raggedright\arraybackslash}m{0.17\linewidth}cccccc@{}}
\morphlab{progress} & \morphlab{0\%} & \morphlab{27\%} & \morphlab{39\%} & \morphlab{63\%} & \morphlab{76\%} & \morphlab{100\%} \\
\morphlab{surface} & \raisebox{-0.5\height}{\includegraphics[width=1.0000\morphtile]{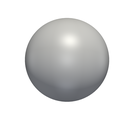}} & \raisebox{-0.5\height}{\includegraphics[width=1.0000\morphtile]{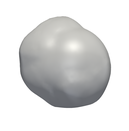}} & \raisebox{-0.5\height}{\includegraphics[width=1.0000\morphtile]{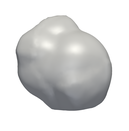}} & \raisebox{-0.5\height}{\includegraphics[width=1.0000\morphtile]{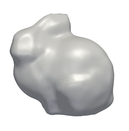}} & \raisebox{-0.5\height}{\includegraphics[width=1.0000\morphtile]{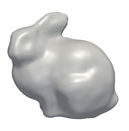}} & \raisebox{-0.5\height}{\includegraphics[width=1.0000\morphtile]{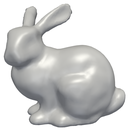}} \\
\morphlab{TetSphere $M{=}1$} & \raisebox{-0.5\height}{\includegraphics[width=1.0000\morphtile]{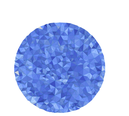}} & \raisebox{-0.5\height}{\includegraphics[width=1.0000\morphtile]{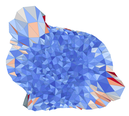}} & \raisebox{-0.5\height}{\includegraphics[width=1.0000\morphtile]{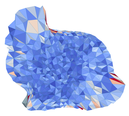}} & \raisebox{-0.5\height}{\includegraphics[width=1.0000\morphtile]{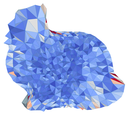}} & \raisebox{-0.5\height}{\includegraphics[width=1.0000\morphtile]{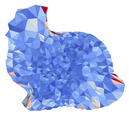}} & \raisebox{-0.5\height}{\includegraphics[width=1.0000\morphtile]{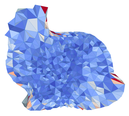}} \\
\morphlab{TetSphere $M{=}6$} & \raisebox{-0.5\height}{\includegraphics[width=1.0000\morphtile]{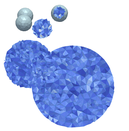}} & \raisebox{-0.5\height}{\includegraphics[width=1.0000\morphtile]{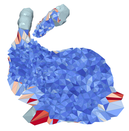}} & \raisebox{-0.5\height}{\includegraphics[width=1.0000\morphtile]{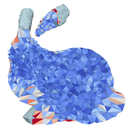}} & \raisebox{-0.5\height}{\includegraphics[width=1.0000\morphtile]{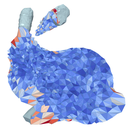}} & \raisebox{-0.5\height}{\includegraphics[width=1.0000\morphtile]{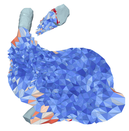}} & \raisebox{-0.5\height}{\includegraphics[width=1.0000\morphtile]{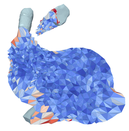}} \\
\morphlab{Fixed\newline connectivity} & \raisebox{-0.5\height}{\includegraphics[width=1.0000\morphtile]{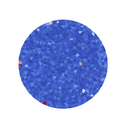}} & \raisebox{-0.5\height}{\includegraphics[width=1.0000\morphtile]{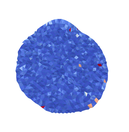}} & \raisebox{-0.5\height}{\includegraphics[width=1.0000\morphtile]{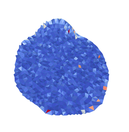}} & \raisebox{-0.5\height}{\includegraphics[width=1.0000\morphtile]{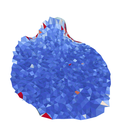}} & \raisebox{-0.5\height}{\includegraphics[width=1.0000\morphtile]{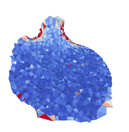}} & \raisebox{-0.5\height}{\includegraphics[width=1.0000\morphtile]{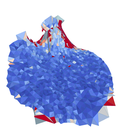}} \\
\morphlab{Ours} & \raisebox{-0.5\height}{\includegraphics[width=1.0000\morphtile]{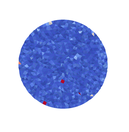}} & \raisebox{-0.5\height}{\includegraphics[width=1.0000\morphtile]{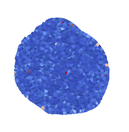}} & \raisebox{-0.5\height}{\includegraphics[width=1.0000\morphtile]{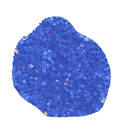}} & \raisebox{-0.5\height}{\includegraphics[width=1.0000\morphtile]{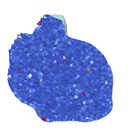}} & \raisebox{-0.5\height}{\includegraphics[width=1.0000\morphtile]{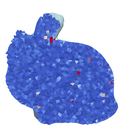}} & \raisebox{-0.5\height}{\includegraphics[width=1.0000\morphtile]{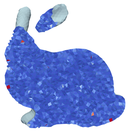}} \\
\morphlab{fTetWild\newline remesh} & \raisebox{-0.5\height}{\includegraphics[width=1.0000\morphtile]{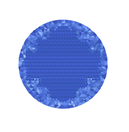}} & \raisebox{-0.5\height}{\includegraphics[width=1.0000\morphtile]{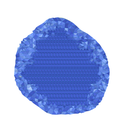}} & \raisebox{-0.5\height}{\includegraphics[width=1.0000\morphtile]{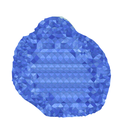}} & \raisebox{-0.5\height}{\includegraphics[width=1.0000\morphtile]{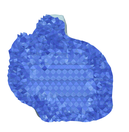}} & \raisebox{-0.5\height}{\includegraphics[width=1.0000\morphtile]{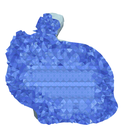}} & \raisebox{-0.5\height}{\includegraphics[width=1.0000\morphtile]{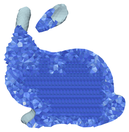}} \\
\end{tabular}
\end{tabular}%
\end{minipage}\hfill
  \begin{minipage}[c]{0.38\textwidth}\includegraphics[width=\linewidth]{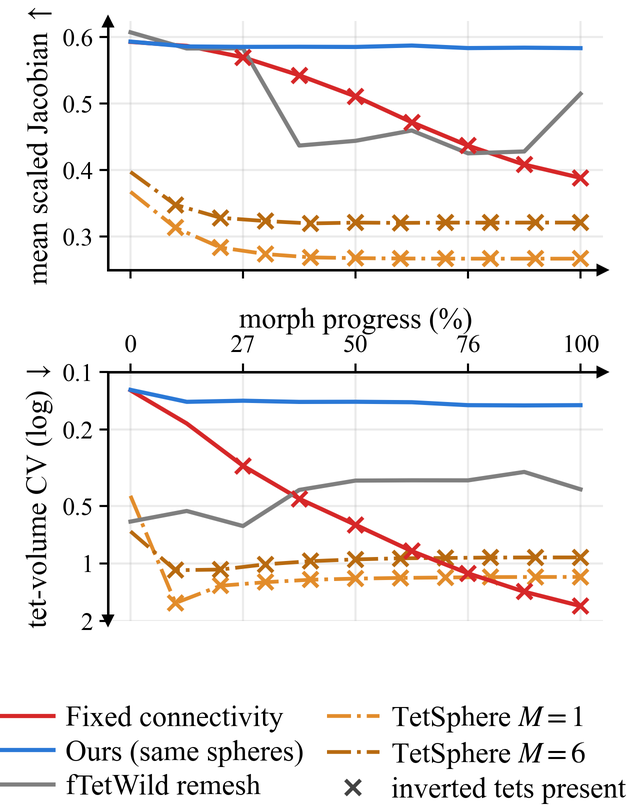}\end{minipage}
  \caption{Sphere-to-bunny morph quality, ${\approx}10$k spheres. Left: surface per step, shared by all but TetSphere; cut views coloured by AMIPS energy, from blue (3, regular) to deep red (8 or more). Right: mean scaled Jacobian and volume CV per step; crosses: inverted elements present.}
  \label{fig:exp-morph}
\end{figure}
\subsection{Identity-preserving morphing}
\label{sec:exp:morph}
We morph a sphere into the bunny along the level-set path of Section~\ref{sec:method:morph} in nine steps, evenly spaced in the resolved fraction of the symmetric-difference volume, with $\phi$ on a $256^3$ grid and 9,941 interior spheres. Ours, fixed connectivity and fTetWild take the same nine surfaces as input and differ in how they treat the interior. Ours re-packs the same spheres and rebuilds the connectivity at each step. Fixed connectivity keeps the step-0 mesh, projects its surface nodes onto each new level set, and moves the interior nodes by harmonic extension of the surface motion. fTetWild meshes each step from scratch with its edge length set for our node count. TetSphere~\citep{guo2024tetsphere} with one ($M{=}1$) or six ($M{=}6$) spheres runs its own 11-step morph of the same pair at about 10k nodes.

Ours has no inverted tetrahedron at any step, and its Vol\,CV stays within 0.124--0.149 (Table~\ref{tab:morph-quality}). Fixed connectivity starts from the same mesh and degrades step by step: inverted tetrahedra appear at step~2 and reach 4.1\% at the bunny. fTetWild has no inverted tetrahedron and three times our volume spread, and TetSphere inverts tetrahedra at every step after the first. Re-packing restores the sphere spacing at every step, while fixed connectivity stretches the source tetrahedra over the new volume. A fresh octree packing at every step, which discards identity, reaches the same Vol\,CV as ours, 0.144 against 0.143, while our interior nodes keep their identities at every step: the sphere colours in Figures~\ref{fig:teaser} and~\ref{fig:overview2d}b follow each sphere from the source to the target. The same procedure carries 64,000 spheres through the morph of Figure~\ref{fig:teaser} without an inverted tetrahedron.

\subsection{Soft robot design study}
\begin{figure}[t]
  \centering
  \resizebox{0.9\textwidth}{!}{
\providecommand{\teaserlab}[1]{{\footnotesize #1}}
\setlength{\tabcolsep}{1pt}\renewcommand{\arraystretch}{0.7}
\begin{tabular}{@{}c@{}}
\raisebox{-0.5\height}{\includegraphics[width=0.1764\textwidth]{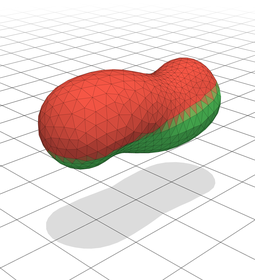}}\begin{tabular}{@{}ccccc@{}}\includegraphics[width=0.1274\textwidth]{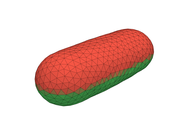} & \includegraphics[width=0.1274\textwidth]{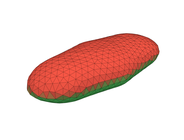} & \includegraphics[width=0.1274\textwidth]{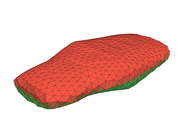} & \includegraphics[width=0.1274\textwidth]{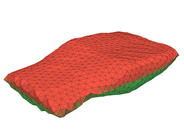} & \includegraphics[width=0.1274\textwidth]{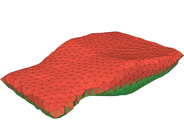} \\ \includegraphics[width=0.1274\textwidth]{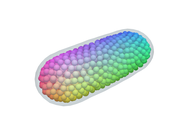} & \includegraphics[width=0.1274\textwidth]{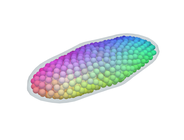} & \includegraphics[width=0.1274\textwidth]{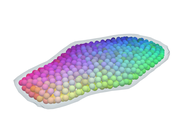} & \includegraphics[width=0.1274\textwidth]{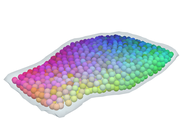} & \includegraphics[width=0.1274\textwidth]{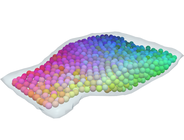} \\ \end{tabular}\raisebox{-0.5\height}{\includegraphics[width=0.1764\textwidth]{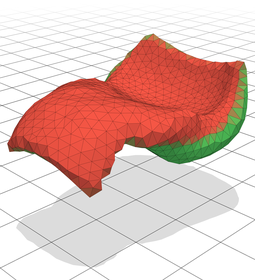}} \\
\teaserlab{(a) Soft robot design iterations: up--down muscle group swimmer} \\[5pt]
\raisebox{-0.5\height}{\includegraphics[width=0.1764\textwidth]{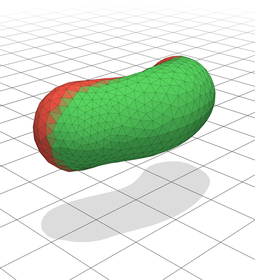}}\begin{tabular}{@{}ccccc@{}}\includegraphics[width=0.1274\textwidth]{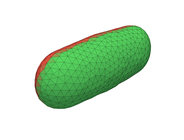} & \includegraphics[width=0.1274\textwidth]{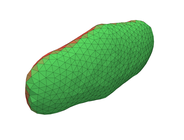} & \includegraphics[width=0.1274\textwidth]{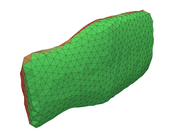} & \includegraphics[width=0.1274\textwidth]{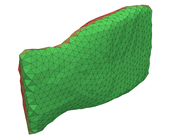} & \includegraphics[width=0.1274\textwidth]{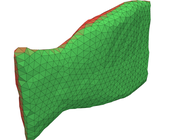} \\ \includegraphics[width=0.1274\textwidth]{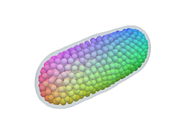} & \includegraphics[width=0.1274\textwidth]{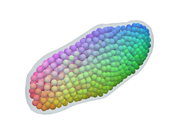} & \includegraphics[width=0.1274\textwidth]{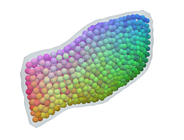} & \includegraphics[width=0.1274\textwidth]{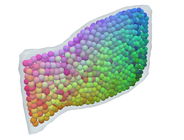} & \includegraphics[width=0.1274\textwidth]{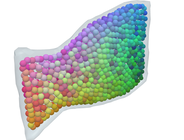} \\ \end{tabular}\raisebox{-0.5\height}{\includegraphics[width=0.1764\textwidth]{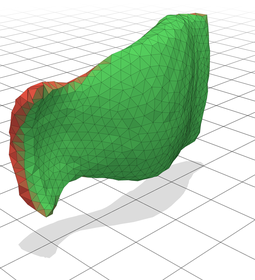}} \\
\teaserlab{(b) Soft robot design iterations: left--right muscle group swimmer} \\[5pt]
\end{tabular}
}
  \caption{Morphology design driven by gradients on surface and interior nodes, 1k nodes: starting capsule mid-stroke, five example iterations, and best design mid-stroke; muscle groups red and green; below each design: its interior spheres, coloured by starting position.}
  \label{fig:design}
  \vspace{6pt}
  {\centering\small\setlength{\tabcolsep}{3pt}
  \captionof{table}{Morph quality, mean$\pm$sd over the morph steps; rows as in Figure~\ref{fig:exp-morph}; Identity: interior node identities kept; $^\dagger$own morph protocol (Section~\ref{sec:exp:morph}).}\label{tab:morph-quality}
  \begin{adjustbox}{max width=\textwidth}
\begin{tabular}{@{}l c c c c c c@{}}
\toprule
Method & Identity & Scaled $J$ $\uparrow$ & \% Inv $\downarrow$ & Min Dih ($^\circ$) $\uparrow$ & AMIPS $\downarrow$ & Vol\,CV $\downarrow$ \\
\midrule
TetSphere $M{=}1$$^\dagger$ & yes & $0.283 \pm 0.030$ & $4.70 \pm 1.85$ & $38.9 \pm 2.9$ & $6.41 \pm 0.95$ & $1.177 \pm 0.261$ \\
TetSphere $M{=}6$$^\dagger$ & yes & $0.331 \pm 0.022$ & \second{$0.44 \pm 0.45$} & $42.1 \pm 1.7$ & $4.51 \pm 0.30$ & $0.950 \pm 0.101$ \\
\midrule
Fixed connectivity & yes & \second{$0.501 \pm 0.078$} & $1.11 \pm 1.59$ & $52.4 \pm 4.4$ & $4.72 \pm 1.94$ & $0.753 \pm 0.552$ \\
Ours & yes & \best{$0.586 \pm 0.003$} & \best{$0.00 \pm 0.00$} & \best{$56.5 \pm 0.3$} & \best{$3.36 \pm 0.02$} & \best{$0.143 \pm 0.008$} \\
fTetWild remesh & no & $0.498 \pm 0.075$ & \best{$0.00 \pm 0.00$} & \second{$54.0 \pm 2.2$} & \second{$3.46 \pm 0.13$} & \second{$0.449 \pm 0.113$} \\
\bottomrule
\end{tabular}
\end{adjustbox}\par}
  \vspace{6pt}
  {\centering\small\setlength{\tabcolsep}{4pt}
  \captionof{table}{Fitness by method, task and mesh size: forward displacement of the centre of mass, averaged over the second half of the rollout, in simulation units; mean$\pm$sd over three seeds.}\label{tab:wave}
  \begin{adjustbox}{max width=\textwidth}
\begin{tabular}{@{}llccccc@{}}
\toprule
environment & scale & ours & ours-surface & remesh-tetgen & remesh-ftw & fixed-conn \\
\midrule
swim, up--down & 1k & \best{$2.469 \pm 0.123$} & \second{$1.395 \pm 0.019$} & $1.300 \pm 0.009$ & $1.254 \pm 0.013$ & $0.535 \pm 0.210$ \\
swim, lateral & 1k & \best{$2.345 \pm 0.031$} & \second{$1.413 \pm 0.027$} & $1.283 \pm 0.031$ & $1.226 \pm 0.023$ & $0.618 \pm 0.239$ \\
loco, up--down & 1k & $1.831 \pm 0.087$ & $1.845 \pm 0.288$ & \second{$1.878 \pm 0.407$} & \best{$1.925 \pm 0.543$} & $1.544 \pm 0.211$ \\
swim, up--down & 8k & \best{$2.488 \pm 0.035$} & \second{$1.761 \pm 0.010$} & $1.700 \pm 0.007$ & $1.687 \pm 0.016$ & $0.600 \pm 0.002$ \\
swim, lateral & 8k & \best{$2.617 \pm 0.079$} & \second{$1.781 \pm 0.012$} & $1.682 \pm 0.009$ & $1.696 \pm 0.026$ & $0.601 \pm 0.001$ \\
loco, up--down & 8k & $1.837 \pm 0.148$ & \second{$1.977 \pm 0.673$} & $1.775 \pm 0.766$ & \best{$2.425 \pm 0.460$} & $1.542 \pm 0.613$ \\
\bottomrule
\end{tabular}
\end{adjustbox}\par}
\end{figure}
\label{sec:exp:robot}
We use the simulator and the morph step to design soft robots (Figure~\ref{fig:design}). A robot is a soft body whose two muscle groups contract in a travelling wave. Every run starts from the same capsule, 1.91 long in simulation units, meshed with about 1k or 8k interior spheres. We test swimming with an up--down and with a left--right muscle split, without gravity, and walking on frictional ground with the up--down split. Fitness is the forward displacement of the centre of mass, averaged over the second half of the rollout; Appendix~\ref{app:designs} lists the settings. Each design iteration runs the mass--spring and Neo-Hookean simulator of Section~\ref{sec:method:sim} for 6\,s, updates the body shape and morphs the mesh into it. Because the simulator is written in JAX, it is differentiable and a rollout with its gradient takes 3.6--6.3\,s on one GPU, so the shape update follows simulator gradients directly.

Five methods share the start shape, the simulator, the gait and the budget. \emph{Ours} takes the gradient over all nodes, interior included, and \emph{ours-surface} only over the surface nodes. \emph{Remesh-TetGen} and \emph{remesh-fTetWild} also use surface-node gradients and rebuild the volume from scratch at every iteration, at the node density of ours. \emph{Fixed connectivity} moves all nodes along a smooth, volume-preserving displacement field built from the node gradients and keeps the initial tetrahedra. Each method runs three seeds in each of the six environment--scale conditions under one protocol, and a run is scored by the best fitness it reaches in 50 iterations, or 100 for walking at 8k.

\begin{figure}[t]
  \centering
{\scriptsize\setlength{\tabcolsep}{0pt}\renewcommand{\arraystretch}{1}%
\begin{tabular}{@{}c@{\hspace{0.004\linewidth}}c@{\hspace{0.004\linewidth}}c@{\hspace{0.004\linewidth}}c@{\hspace{0.004\linewidth}}c@{\hspace{0.004\linewidth}}c@{\hspace{0.004\linewidth}}c@{}}
\makebox[0.03\linewidth]{} & $t{=}0.0$\,s & $t{=}1.2$\,s & $t{=}2.4$\,s & $t{=}3.6$\,s & $t{=}4.8$\,s & $t{=}6.0$\,s \\[1pt]
\makebox[0.03\linewidth]{\rotatebox[origin=c]{90}{\textcolor[HTML]{0072B2}{fixed wave}}} & \raisebox{-.5\height}{\includegraphics[width=0.1575\linewidth]{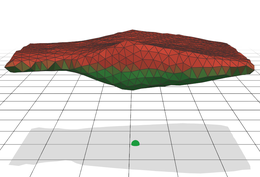}} & \raisebox{-.5\height}{\includegraphics[width=0.1575\linewidth]{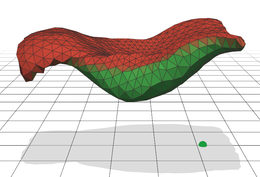}} & \raisebox{-.5\height}{\includegraphics[width=0.1575\linewidth]{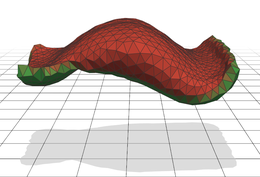}} & \raisebox{-.5\height}{\includegraphics[width=0.1575\linewidth]{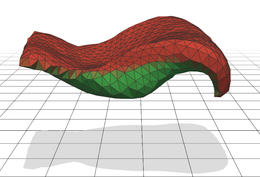}} & \raisebox{-.5\height}{\includegraphics[width=0.1575\linewidth]{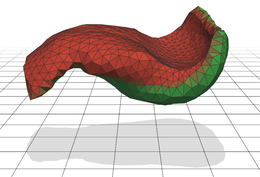}} & \raisebox{-.5\height}{\includegraphics[width=0.1575\linewidth]{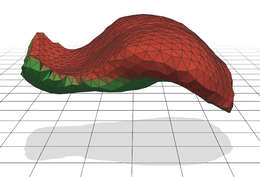}} \\
 & $\Delta x{=}{+}0.00$ & $\Delta x{=}{+}0.70$ & $\Delta x{=}{+}1.73$ & $\Delta x{=}{+}2.76$ & $\Delta x{=}{+}3.79$ & $\Delta x{=}{+}4.84$ \\[2pt]
\makebox[0.03\linewidth]{\rotatebox[origin=c]{90}{\textcolor[HTML]{D55E00}{learned controller}}} & \raisebox{-.5\height}{\includegraphics[width=0.1575\linewidth]{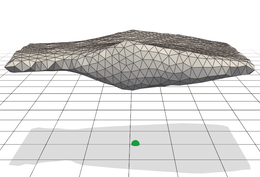}} & \raisebox{-.5\height}{\includegraphics[width=0.1575\linewidth]{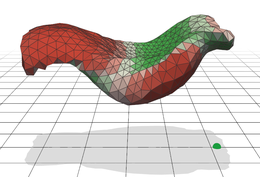}} & \raisebox{-.5\height}{\includegraphics[width=0.1575\linewidth]{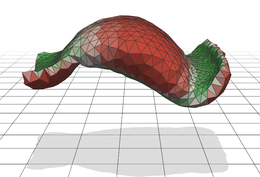}} & \raisebox{-.5\height}{\includegraphics[width=0.1575\linewidth]{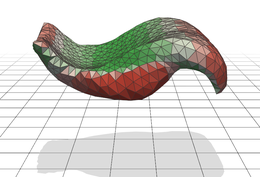}} & \raisebox{-.5\height}{\includegraphics[width=0.1575\linewidth]{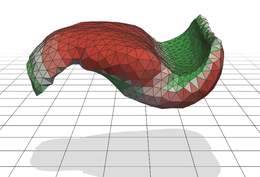}} & \raisebox{-.5\height}{\includegraphics[width=0.1575\linewidth]{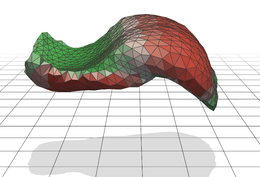}} \\
 & $\Delta x{=}{+}0.00$ & $\Delta x{=}{+}0.85$ & $\Delta x{=}{+}2.08$ & $\Delta x{=}{+}3.32$ & $\Delta x{=}{+}4.57$ & $\Delta x{=}{+}5.81$ \\
\end{tabular}}%

  \caption{Fixed-wave and learned-controller swimmers: one design at the same time points. Top: fixed travelling wave, coloured by muscle group; bottom: learned controller, coloured by muscle activation; $\Delta x$: forward displacement at time $t$.}
  \label{fig:frames}
  \vspace{4pt}
  \includegraphics[width=0.60\textwidth]{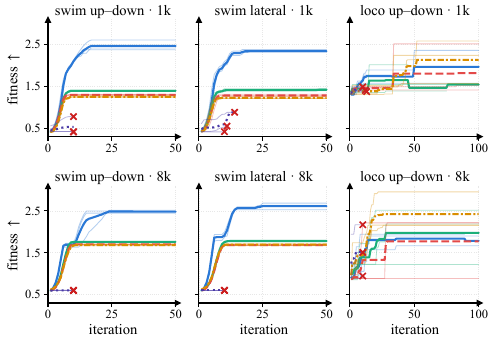}\hfill
  \includegraphics[width=0.38\textwidth]{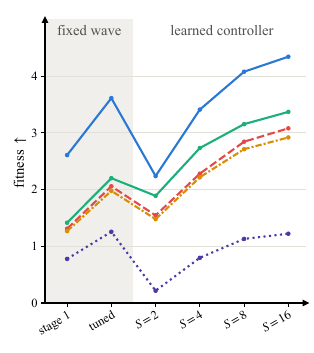}\\[2pt]
  \includegraphics[width=\textwidth]{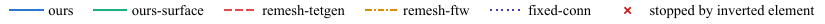}
  \caption{Shape-optimization curves and controller tests. Left: fitness per design iteration at 1k (top) and 8k nodes (bottom), walking at 1k from a separate 100-iteration rerun; thin lines: single runs; thick lines: mean of active runs. Right: the best 1k up--down designs at six test points: design-stage wave, tuned wave, and learned controllers of size $S{=}2$--$16$.}
  \label{fig:curves}
\end{figure}
\newcommand{\VoxelBlock}{%
\begin{wrapfigure}[17]{r}{\dimexpr0.5\textwidth-\columnsep\relax}
  \centering
{\scriptsize\setlength{\tabcolsep}{0pt}\renewcommand{\arraystretch}{1}%
\begin{tabular}{@{}c@{\hspace{0.019\linewidth}}c@{}}
\begin{tabular}[b]{@{}c@{}}ours: smooth surface\\[1pt]
   {\footnotesize\bfseries +2.563}\,sim.\ units\\ (1.13 body lengths)\end{tabular}
 & \begin{tabular}[b]{@{}c@{}}voxel staircase\\[1pt]
   {\footnotesize\bfseries +1.755}\,sim.\ units\\ (0.77 body lengths)\end{tabular} \\[1pt]
\raisebox{-.5\height}{\includegraphics[width=0.46\linewidth]{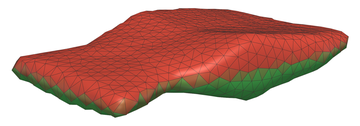}}
 & \raisebox{-.5\height}{\includegraphics[width=0.46\linewidth]{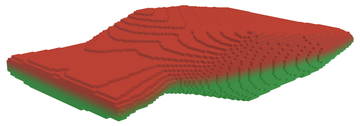}} \\ \noalign{\vspace{3pt}}
\raisebox{-.5\height}{\includegraphics[width=0.46\linewidth]{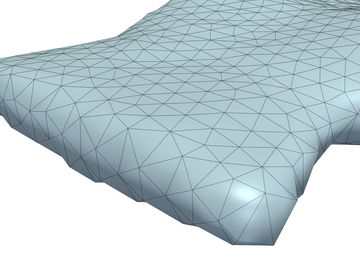}}
 & \raisebox{-.5\height}{\includegraphics[width=0.46\linewidth]{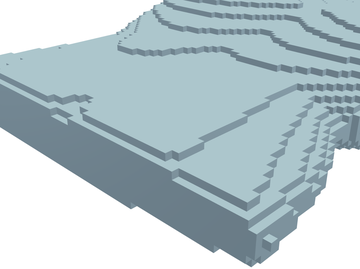}} \\
\end{tabular}}%

  \caption{Mesh and voxel design representations: one swimming design with a smooth surface (left) and its voxel version on the $128^3$ design grid (right), with their fitness.}
  \label{fig:voxel}
\end{wrapfigure}
}
In all four swimming conditions, ours reaches the highest fitness and exceeds ours-surface by 0.73--1.07, and the seed ranges of the two methods are disjoint (Table~\ref{tab:wave}). Ours and ours-surface share every step except the gradient support, so this margin measures what the interior gradients contribute. The remeshing baselines finish close to ours-surface, and fixed connectivity stops near iteration 10, when an element inverts. Section~\ref{sec:conclusion} discusses walking.

Both swimmers turn the capsule into a plate that faces the direction of bending (Figure~\ref{fig:design}). In the drag model, a face moving along its normal meets six times the resistance of a face sliding along itself, so broad faces that sweep along their normals produce thrust, and the bending that moves them depends on how material is distributed through the whole body. Gradients on interior nodes measure how moving material inside the body changes the fitness; surface-node gradients measure how moving the body outline changes it.

\begin{wrapfigure}[17]{r}{\dimexpr0.5\textwidth-\columnsep\relax}
  \centering
{\scriptsize\setlength{\tabcolsep}{0pt}\renewcommand{\arraystretch}{1}%
\begin{tabular}{@{}c@{\hspace{0.019\linewidth}}c@{}}
\begin{tabular}[b]{@{}c@{}}ours: smooth surface\\[1pt]
   {\footnotesize\bfseries +2.563}\,sim.\ units\\ (1.13 body lengths)\end{tabular}
 & \begin{tabular}[b]{@{}c@{}}voxel staircase\\[1pt]
   {\footnotesize\bfseries +1.755}\,sim.\ units\\ (0.77 body lengths)\end{tabular} \\[1pt]
\raisebox{-.5\height}{\includegraphics[width=0.46\linewidth]{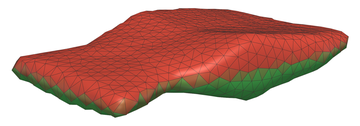}}
 & \raisebox{-.5\height}{\includegraphics[width=0.46\linewidth]{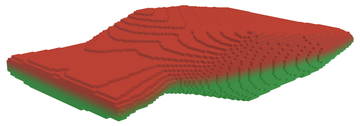}} \\ \noalign{\vspace{3pt}}
\raisebox{-.5\height}{\includegraphics[width=0.46\linewidth]{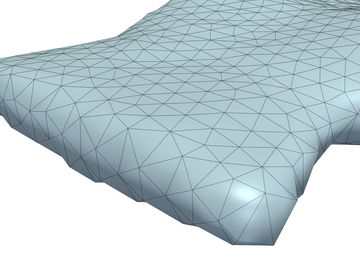}}
 & \raisebox{-.5\height}{\includegraphics[width=0.46\linewidth]{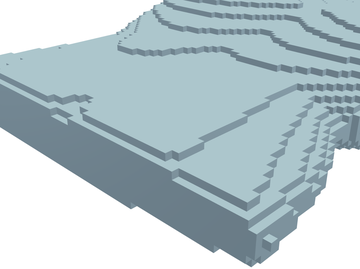}} \\
\end{tabular}}%

  \caption{Mesh and voxel design representations: one swimming design with a smooth surface (left) and its voxel version on the $128^3$ design grid (right), with their fitness.}
  \label{fig:voxel}
\end{wrapfigure}
A design ranked under one fixed gait may lose its rank under a stronger controller, so we train learned controllers on the best up--down design of each method at 1k: a shared multilayer perceptron maps clock signals and the rest position of each muscle block to its activation and is trained through the simulator (Appendix~\ref{app:designs}).

Each controller starts from a least-squares fit to the travelling wave, and each test point is scored by the median fitness over three start phases. The order ours $>$ ours-surface $>$ remesh-TetGen $>$ remesh-fTetWild $>$ fixed connectivity holds at all six test points of Figure~\ref{fig:curves}, and the ratio of ours to ours-surface falls from 1.84 after design to 1.29 with the largest controller. In Figure~\ref{fig:frames}, the learned controller carries our design farther than the fixed wave. With the largest controller, ours reaches a fitness of 4.34, 1.20$\times$ its fitness under the tuned wave, and four of the five methods exceed their tuned-wave fitness. Played back for 6\,s, the best 1k up--down design of ours travels 3.50 under the design-stage wave and 5.81 under the learned controller, against 1.69 and 3.89 for remesh-fTetWild (Figure~\ref{fig:app-swim}). The final designs show where the interior gradients lead (Figure~\ref{fig:app-designs}): in up--down swimming, ours flattens the capsule to a height-to-width ratio of 0.30--0.31, while the three surface-gradient methods reach 0.65--0.74; in left--right swimming, ours raises the ratio to 3.51--3.97, against 1.33--1.54; fixed connectivity keeps 0.82--1.27, close to the capsule. Interior gradients thus carry the body towards the broad plates that the drag model rewards.

To evaluate voxel versions of our swimming designs, we replace their tetrahedral body by a voxel body with the same gait and objective, at a resolution matched to our surface and on the $128^3$ design grid (Figure~\ref{fig:voxel}). The voxel bodies reach 32--63\% lower fitness than the smooth designs at both resolutions (Appendix~\ref{app:voxel}).

\section{Discussion}
\label{sec:conclusion}
\textbf{Limitations and future work.} (i)~\emph{Computational cost.} Our pipeline needs a GPU and meshes a shape more slowly than TetGen, and thin boundary tetrahedra shorten the stable time step of explicit integration compared with TetGen and fTetWild meshes. (ii)~\emph{Identity.} To keep the surface triangles well shaped, every step re-triangulates the surface before CDT, so surface nodes are not tracked and only interior nodes persist; in the design study the optimizer keeps its momentum on the FFD lattice, so identity serves correspondence without a direct path to fitness, and element quality varies together with identity across the baselines. (iii)~\emph{Experimental scope.} Walking remains inconclusive, because ground contact makes its gradient non-smooth, and GPU atomics make repeated gradient evaluations differ slightly (Appendix~\ref{app:simspeed}); a few conditions differ in budget or protocol: walking at 8k, four runs cut short by CDT failures, one controller run per condition, fTetWild overshooting its node target at the first morph steps, TetSphere's own morph, and the voxel discretization; the simulator uses simulation units with equal node masses, so body mass grows with node count. In future work, a material field stored on the persistent nodes and interpolated to each new mesh would give identity a direct role in design.

\textbf{Conclusion.} \method{} builds tetrahedral meshes from physical sphere packings: spheres flow and jam inside a re-triangulated surface, and CDT connects them into a conforming mesh with more uniform element volumes than TetGen and fTetWild. Because the same spheres re-settle inside every new shape, interior nodes keep their identities through large shape changes while the connectivity follows the shape, and the meshes stay free of inverted elements where fixed-connectivity and TetSphere meshes invert. A differentiable GPU simulator couples these meshes with morphology design, where gradients on interior nodes raise swimming fitness by 0.73--1.07 over surface-only gradients, and our designs reach 1.5--2.7$\times$ the fitness of their voxel versions. The same meshes also support compliance design with linear finite elements (Appendix~\ref{app:fea}); together, these results make sphere packing a practical volumetric mesh representation for gradient-based shape optimization.

\subsection*{Reproducibility statement}
Our code, under review with this paper, covers meshing, morphing, simulation and the design loop. Section~\ref{sec:method} and Appendix~\ref{app:designs} give the settings of packing, morphing, simulation and design; Appendix~\ref{app:simspeed} reports the simulator's speed and gradient checks; Appendices~\ref{app:static}, \ref{app:voxel} and~\ref{app:fea} give the full static, voxel and finite-element results. The design study runs three seeds per method and condition. Experiments run on NVIDIA RTX 5090 GPUs under Ubuntu.

\subsection*{AI use statement}
The author used an LLM for language editing, formatting, and as a search tool in the literature review, in parts of this paper and its appendix. All scientific ideas, claims, decisions, and the selection and interpretation of the cited literature are those of the researcher.

\bibliography{references,related_extra}
\bibliographystyle{iclr2027_conference}

\clearpage
\appendix
\renewcommand{\thefigure}{A\arabic{figure}}\setcounter{figure}{0}
\renewcommand{\thetable}{A\arabic{table}}\setcounter{table}{0}

\section{Soft-robot design representations}
\label{app:repr}
\begin{table}[h]
  \centering\small\setlength{\tabcolsep}{4pt}
  \caption{Body representations in soft-robot design. Free-form: \checkmark{} any closed level-set surface; $\sim$ shape made of grid cells or particles; $\times$ primitives or a template. Diff.: simulator gradients.}
  \label{tab:app-review}
  \begin{adjustbox}{max width=\textwidth}
\begin{tabular}{@{}l l l c l l c@{}}
\toprule
Work & Representation & Surface & Free-form & Task & Compute & Diff. \\
\midrule
Sims~\citeyearpar{sims1994evolving} & primitives & blocky & $\times$ & diverse & CM-5 & $\times$ \\
Cheney et al.~\citeyearpar{cheney2013unshackling} & voxels ($10^3$) & blocky & $\sim$ & forward locomotion & VoxCAD / CPU & $\times$ \\
EvoGym~\citeyearpar{bhatia2021evogym} & voxels ($5{\times}5$) & blocky & $\sim$ & diverse (2D) & C++ / CPU & $\times$ \\
DiffAqua~\citeyearpar{ma2021diffaqua} & voxel grid & blocky & $\times$ & forward swim & DiffPD / CPU & \checkmark \\
SoftZoo~\citeyearpar{wang2023softzoo} & MPM particles & particle & $\sim$ & directional locomotion & Taichi / GPU & \checkmark \\
\midrule
Ours & conforming tet mesh & smooth & \checkmark & swimming, walking & JAX / GPU & \checkmark \\
\bottomrule
\end{tabular}
\end{adjustbox}
\end{table}
Table~\ref{tab:app-review} compares the body representation of our design pipeline with earlier soft-robot design systems. Primitive and voxel bodies have blocky surfaces, and particle bodies have no explicit surface. Each of these systems restricts the body to primitives, a template or a grid, and two of them provide simulator gradients. Our body is a conforming tetrahedral mesh with a smooth surface, built from a closed level-set surface, and the simulator provides gradients with respect to its nodes.

\section{Static meshing on nine shapes}
\label{app:static}
\begin{figure}[p]
  \centering
  \input{figure/app_gallery_grid.tex}
  \caption{Seven shapes at 64k interior spheres. Rows per block: packed spheres; surface mesh; cut view coloured by AMIPS energy: regular \raisebox{-0.15ex}{\includegraphics[width=3em,height=1.1ex]{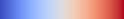}} distorted.}
  \label{fig:app-gallery}
\end{figure}

Figure~\ref{fig:app-gallery} shows seven of the nine shapes of Section~\ref{sec:exp:static}, and Table~\ref{tab:app-static} lists the statistics of all nine. Our Vol\,CV lies in 0.127--0.207, against 0.293--0.308 for TetGen and 0.310--0.540 for fTetWild, and no mesher produces an inverted element. Our mean AMIPS energy is 3.32--3.47, the lowest or second lowest on every shape. Our pipeline takes 35.9--123.4\,s per shape, TetGen under one second and fTetWild 6.7--1340\,s. The I-beam gives our highest Vol\,CV and the dragon our longest run.
\begin{table}[h]
  \centering\small\setlength{\tabcolsep}{3pt}
  \caption{Static mesh statistics at 64k interior spheres for the nine shapes of Figure~\ref{fig:octree}. Nodes and Tets: our meshes; shading per shape; no inverted elements for any method.}
  \label{tab:app-static}
  \begin{adjustbox}{max width=\textwidth}
\begin{tabular}{@{}l r r ccc ccc rrr@{}}
\toprule
 & & & \multicolumn{3}{c}{Vol\,CV $\downarrow$} & \multicolumn{3}{c}{AMIPS $\downarrow$} & \multicolumn{3}{c}{Wall (s)} \\
\cmidrule(lr){4-6}\cmidrule(lr){7-9}\cmidrule(l){10-12}
Shape & Nodes & Tets & Ours & TetGen & fTetWild & Ours & TetGen & fTetWild & Ours & TetGen & fTetWild \\
\midrule
Sphere & 74,242 & 415,041 & \best{0.127} & \second{0.297} & 0.441 & \best{3.33} & 3.81 & \second{3.41} & 36.8 & \best{0.4} & \second{16.2} \\
Cube & 76,636 & 421,981 & \best{0.130} & \second{0.298} & 0.310 & \second{3.32} & 3.81 & \best{3.31} & 35.9 & \best{0.5} & \second{7.4} \\
Beam & 87,379 & 455,294 & \best{0.155} & \second{0.293} & 0.324 & \best{3.37} & 3.79 & \second{3.42} & \second{41.2} & \best{0.6} & 55.6 \\
I-beam & 104,083 & 511,809 & \best{0.207} & \second{0.308} & 0.355 & \best{3.47} & 3.81 & \second{3.49} & 44.6 & \best{0.8} & \second{6.7} \\
Spot & 78,577 & 427,679 & \best{0.134} & \second{0.296} & 0.332 & \best{3.33} & 3.80 & \second{3.49} & \second{36.4} & \best{0.5} & 50.3 \\
Bob & 79,513 & 430,960 & \best{0.140} & \second{0.300} & 0.329 & \best{3.34} & 3.82 & \second{3.49} & \second{37.0} & \best{0.5} & 52.0 \\
Bunny & 77,726 & 425,192 & \best{0.134} & \second{0.296} & 0.540 & \second{3.33} & 3.81 & \best{3.33} & \second{36.9} & \best{0.5} & 77.2 \\
Armadillo & 82,870 & 437,233 & \best{0.149} & \second{0.298} & 0.342 & \best{3.35} & 3.81 & \second{3.41} & \second{41.0} & \best{0.6} & 1340.0 \\
Dragon & 89,507 & 455,341 & \best{0.167} & \second{0.300} & 0.339 & \best{3.39} & 3.81 & \second{3.42} & \second{123.4} & \best{0.7} & 887.2 \\
\bottomrule
\end{tabular}
\end{adjustbox}
\end{table}

\section{Simulator speed}
\label{app:simspeed}
\begin{table}[h]
  \centering\small\setlength{\tabcolsep}{5pt}
  \caption{Simulator speed on one RTX 5090 for a 6\,s rollout (24,000 steps); median of five calls after a warm-up, compilation excluded; memory: peak GPU memory, forward / with gradient.}
  \label{tab:app-simspeed}
  \begin{tabular}{@{}l r r r r r r@{}}
  \toprule
  Mesh & Nodes & Tets & Forward (s) & Steps/s & With gradient (s) & Memory (MiB) \\
  \midrule
  Capsule, 1k & 1,680 & 7,954 & 0.40 & 60,600 & 3.56 & 654 / 1,678 \\
  Capsule, 8k & 10,768 & 56,438 & 0.79 & 30,200 & 6.28 & 656 / 8,852 \\
  Bunny, 64k & 77,726 & 425,192 & 2.27 & 10,600 & -- & 760 / -- \\
  \bottomrule
  \end{tabular}
\end{table}
Table~\ref{tab:app-simspeed} times the simulator of Section~\ref{sec:method:sim} in the swimming setting of Section~\ref{sec:exp:robot} on one NVIDIA RTX 5090, with JAX 0.9 in single precision. A 6\,s rollout of the 1k capsule runs in 0.40\,s, 15 times faster than real time, and its gradient with respect to all rest positions takes 3.56\,s, 9.0 times the forward cost. The 8k capsule takes 0.79\,s forward and 6.28\,s with its gradient, 7.9 times the forward cost. The forward rollout of the 64k bunny takes 2.27\,s; its gradient was not run, because the compiled program was estimated to need more than 82\% of the card's 32\,GB. The benchmark uses checkpoint blocks of 500 steps, and compilation takes 2.8--13.4\,s.

\textbf{Gradient checks.} At a fixed mesh, the adjoint gradient of a rollout with respect to element stiffness agrees with central finite differences to $5.5\times10^{-9}$ in relative terms without ground contact. With contact, the two depart as the difference step shrinks, with relative errors of 0.19, 0.55 and 0.97 at $\epsilon = 10^{-2}, 10^{-3}, 10^{-4}$, so the walking objective is non-smooth. Repeated gradient evaluations on one GPU differ by $2.3\times10^{-3}$ in relative terms, because unordered scatter-adds run as atomics.

\section{Robot designs}
\label{app:designs}
\textbf{Gait.} A plane splits the muscle edges into two groups, which contract in antiphase: the rest length of a muscle edge is scaled by $a_e(t) = 1 \pm A\sin\big(2\pi(ft - \xi_e)\big)$ with $f = 2$\,Hz, where the phase $\xi_e$ grows by one period from head to tail; $A = 0.18$ for swimming and $A = 0.12$ for walking. Swimming uses the drag model of Section~\ref{sec:method:sim} without gravity, and walking uses frictional ground.

\textbf{Setup.} All quantities are in simulation units. The simulator steps at 0.25\,ms, 24,000 steps per 6\,s rollout, with node mass 0.0125, edge stiffness 4,000, edge damping 3 and Neo-Hookean modulus 4,000; swimming uses normal and tangential drag coefficients of 480 and 80. The design objective adds penalties on lateral drift and body orientation to the fitness. Each gradient averages eight rollouts of perturbed shapes, four antithetic pairs with a perturbation RMS of 0.03. The FFD lattice spacing is 8 sphere spacings for swimming and 12 for walking at 1k, and 16 and 24 at 8k, where the spheres lie half as far apart; the lattice has $5^3$ control points for swimming and $4^3$ for walking.
Four of the 90 runs ended early on a CDT failure after their best fitness had plateaued, and each is scored by the best fitness it reached.

\textbf{Controllers.} Following the clock inputs of ELDiR~\citep{strgar2024evolution}, a shared multilayer perceptron maps eight clock signals and the rest position of each of $2S$ muscle blocks to that block's activation, for $S \in \{2,4,8,16\}$. It reads no simulation state and, as in ChainQueen~\citep{hu2019chainqueen}, is trained by gradient descent through the simulator, with one training run and the same budget for each method and size. The least-squares initialization fits the travelling wave with normalized error 0.725 at $S{=}2$ and 0.114 at $S{=}16$. The six test points of Figure~\ref{fig:curves} are the design stage, the fixed gait with tuned amplitude and period, and the four controller sizes; with the tuned gait, ours reaches 1.64$\times$ the fitness of ours-surface.

\begin{figure}[t]
  \centering
{\scriptsize\setlength{\tabcolsep}{0pt}\renewcommand{\arraystretch}{1}%
\begin{tabular}{@{}c@{\hspace{0.004\linewidth}}c@{\hspace{0.004\linewidth}}c@{\hspace{0.004\linewidth}}c@{\hspace{0.004\linewidth}}c@{\hspace{0.004\linewidth}}c@{}}
\makebox[0.03\linewidth]{} & Ours & Ours-surface & Remesh-TetGen & Remesh-fTetWild & Fixed conn. \\[1pt]
\multicolumn{6}{@{}l}{\small 1k interior spheres} \\[1pt]
\makebox[0.03\linewidth]{\rotatebox[origin=c]{90}{swim, up--down}} & \raisebox{-.5\height}{\includegraphics[width=0.190\linewidth]{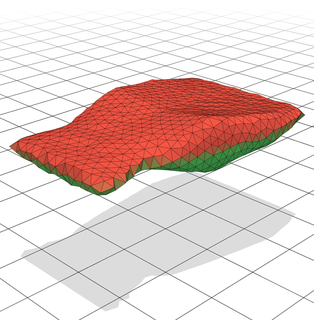}} & \raisebox{-.5\height}{\includegraphics[width=0.190\linewidth]{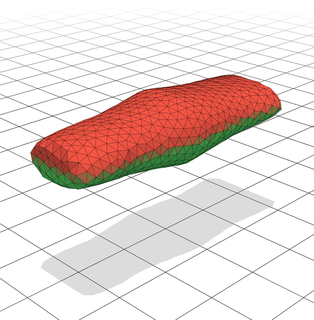}} & \raisebox{-.5\height}{\includegraphics[width=0.190\linewidth]{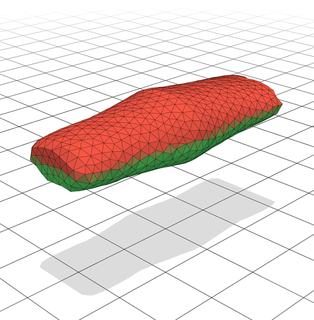}} & \raisebox{-.5\height}{\includegraphics[width=0.190\linewidth]{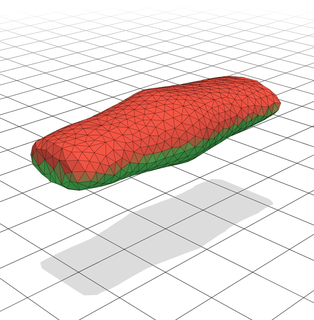}} & \raisebox{-.5\height}{\includegraphics[width=0.190\linewidth]{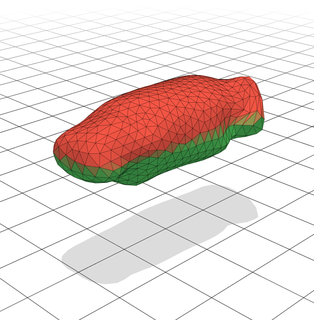}} \\
\makebox[0.03\linewidth]{\rotatebox[origin=c]{90}{swim, left--right}} & \raisebox{-.5\height}{\includegraphics[width=0.190\linewidth]{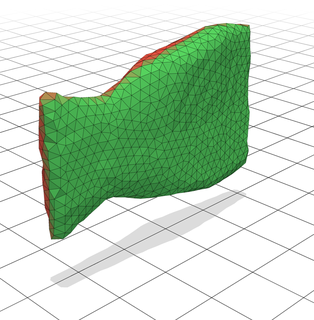}} & \raisebox{-.5\height}{\includegraphics[width=0.190\linewidth]{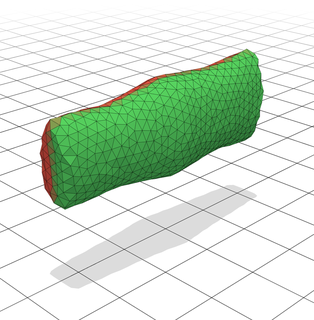}} & \raisebox{-.5\height}{\includegraphics[width=0.190\linewidth]{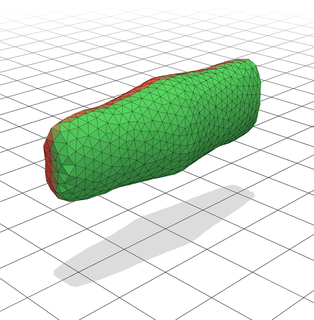}} & \raisebox{-.5\height}{\includegraphics[width=0.190\linewidth]{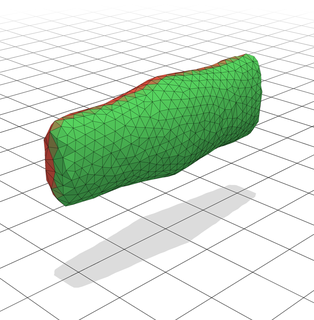}} & \raisebox{-.5\height}{\includegraphics[width=0.190\linewidth]{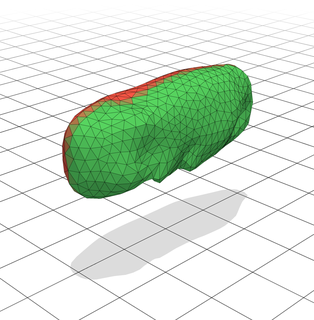}} \\
\noalign{\vspace{4pt}}
\multicolumn{6}{@{}l}{\small 8k interior spheres} \\[1pt]
\makebox[0.03\linewidth]{\rotatebox[origin=c]{90}{swim, up--down}} & \raisebox{-.5\height}{\includegraphics[width=0.190\linewidth]{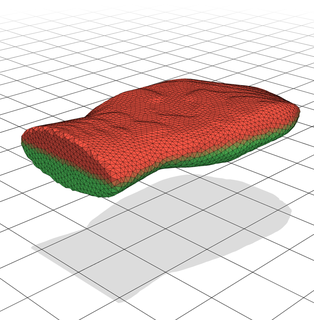}} & \raisebox{-.5\height}{\includegraphics[width=0.190\linewidth]{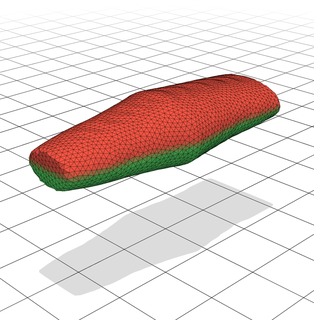}} & \raisebox{-.5\height}{\includegraphics[width=0.190\linewidth]{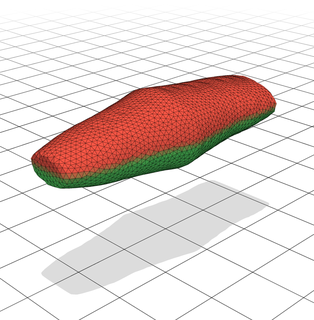}} & \raisebox{-.5\height}{\includegraphics[width=0.190\linewidth]{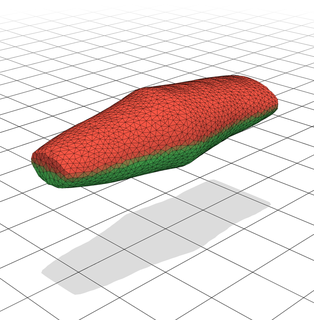}} & \raisebox{-.5\height}{\includegraphics[width=0.190\linewidth]{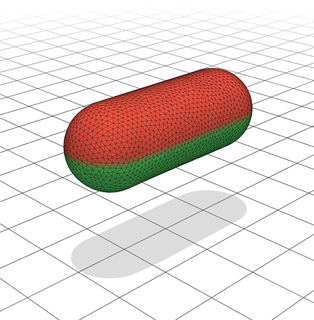}} \\
\makebox[0.03\linewidth]{\rotatebox[origin=c]{90}{swim, left--right}} & \raisebox{-.5\height}{\includegraphics[width=0.190\linewidth]{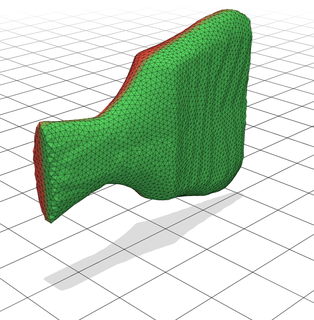}} & \raisebox{-.5\height}{\includegraphics[width=0.190\linewidth]{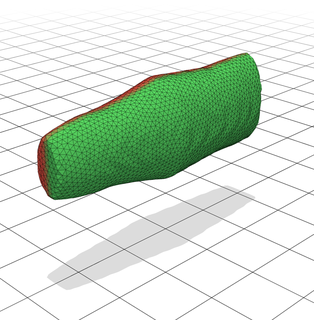}} & \raisebox{-.5\height}{\includegraphics[width=0.190\linewidth]{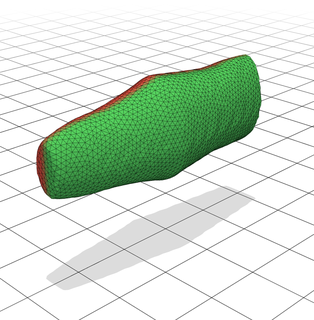}} & \raisebox{-.5\height}{\includegraphics[width=0.190\linewidth]{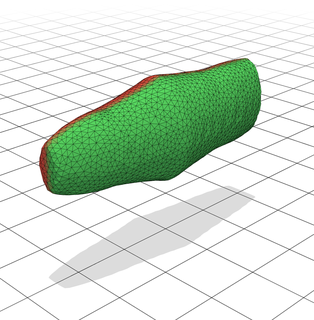}} & \raisebox{-.5\height}{\includegraphics[width=0.190\linewidth]{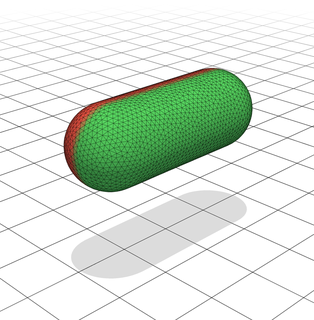}} \\
\end{tabular}}%

  \caption{Final design of the best seed in every swimming condition of Table~\ref{tab:wave}, at rest, coloured by muscle group; columns: methods; rows: up--down and left--right swimming; top block 1k, bottom block 8k interior spheres; same scale in all panels.}
  \label{fig:app-designs}
\end{figure}
Figure~\ref{fig:app-designs} shows the final design of the best seed in every swimming condition, grown from the same capsule, 1.91 long with a $0.72 \times 0.72$ cross-section; Section~\ref{sec:exp:robot} compares their shapes.

\begin{figure}[t]
  \centering
{\scriptsize\setlength{\tabcolsep}{0pt}\renewcommand{\arraystretch}{1}%
\begin{tabular}{@{}c@{\hspace{0.004\linewidth}}c@{\hspace{0.004\linewidth}}c@{\hspace{0.004\linewidth}}c@{\hspace{0.004\linewidth}}c@{\hspace{0.004\linewidth}}c@{\hspace{0.004\linewidth}}c@{}}
\makebox[0.03\linewidth]{} & $t{=}0.0$\,s & $t{=}1.2$\,s & $t{=}2.4$\,s & $t{=}3.6$\,s & $t{=}4.8$\,s & $t{=}6.0$\,s \\[1pt]
\makebox[0.03\linewidth]{\rotatebox[origin=c]{90}{\textcolor[HTML]{0072B2}{fTetWild, wave}}} & \raisebox{-.5\height}{\includegraphics[width=0.1575\linewidth]{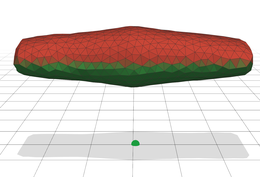}} & \raisebox{-.5\height}{\includegraphics[width=0.1575\linewidth]{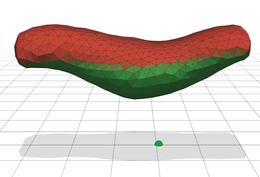}} & \raisebox{-.5\height}{\includegraphics[width=0.1575\linewidth]{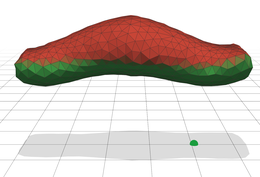}} & \raisebox{-.5\height}{\includegraphics[width=0.1575\linewidth]{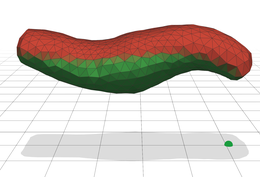}} & \raisebox{-.5\height}{\includegraphics[width=0.1575\linewidth]{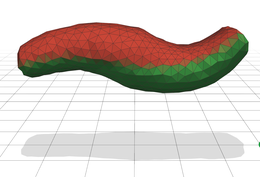}} & \raisebox{-.5\height}{\includegraphics[width=0.1575\linewidth]{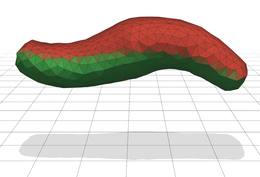}} \\
 & $\Delta x{=}{+}0.00$ & $\Delta x{=}{+}0.24$ & $\Delta x{=}{+}0.61$ & $\Delta x{=}{+}0.97$ & $\Delta x{=}{+}1.33$ & $\Delta x{=}{+}1.69$ \\[2pt]
\makebox[0.03\linewidth]{\rotatebox[origin=c]{90}{\textcolor[HTML]{0072B2}{ours, wave}}} & \raisebox{-.5\height}{\includegraphics[width=0.1575\linewidth]{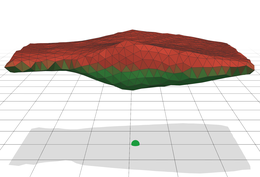}} & \raisebox{-.5\height}{\includegraphics[width=0.1575\linewidth]{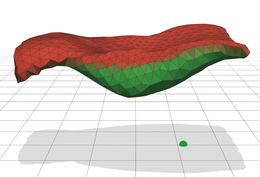}} & \raisebox{-.5\height}{\includegraphics[width=0.1575\linewidth]{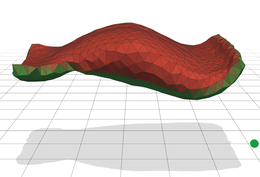}} & \raisebox{-.5\height}{\includegraphics[width=0.1575\linewidth]{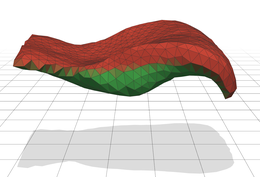}} & \raisebox{-.5\height}{\includegraphics[width=0.1575\linewidth]{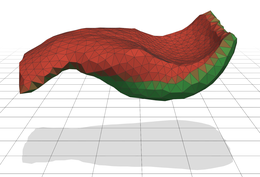}} & \raisebox{-.5\height}{\includegraphics[width=0.1575\linewidth]{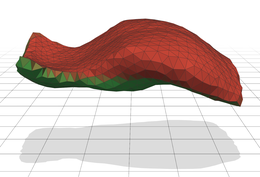}} \\
 & $\Delta x{=}{+}0.00$ & $\Delta x{=}{+}0.50$ & $\Delta x{=}{+}1.24$ & $\Delta x{=}{+}1.99$ & $\Delta x{=}{+}2.74$ & $\Delta x{=}{+}3.50$ \\[2pt]
\makebox[0.03\linewidth]{\rotatebox[origin=c]{90}{\textcolor[HTML]{D55E00}{fTetWild, $S{=}16$}}} & \raisebox{-.5\height}{\includegraphics[width=0.1575\linewidth]{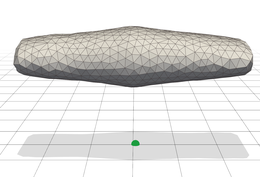}} & \raisebox{-.5\height}{\includegraphics[width=0.1575\linewidth]{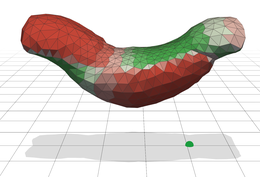}} & \raisebox{-.5\height}{\includegraphics[width=0.1575\linewidth]{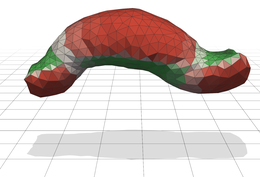}} & \raisebox{-.5\height}{\includegraphics[width=0.1575\linewidth]{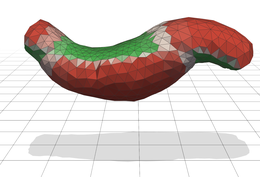}} & \raisebox{-.5\height}{\includegraphics[width=0.1575\linewidth]{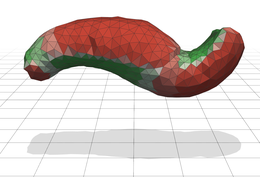}} & \raisebox{-.5\height}{\includegraphics[width=0.1575\linewidth]{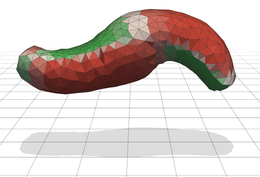}} \\
 & $\Delta x{=}{+}0.00$ & $\Delta x{=}{+}0.56$ & $\Delta x{=}{+}1.41$ & $\Delta x{=}{+}2.25$ & $\Delta x{=}{+}3.07$ & $\Delta x{=}{+}3.89$ \\[2pt]
\makebox[0.03\linewidth]{\rotatebox[origin=c]{90}{\textcolor[HTML]{D55E00}{ours, $S{=}16$}}} & \raisebox{-.5\height}{\includegraphics[width=0.1575\linewidth]{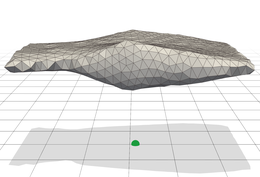}} & \raisebox{-.5\height}{\includegraphics[width=0.1575\linewidth]{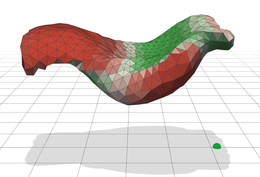}} & \raisebox{-.5\height}{\includegraphics[width=0.1575\linewidth]{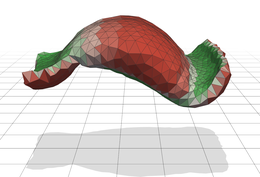}} & \raisebox{-.5\height}{\includegraphics[width=0.1575\linewidth]{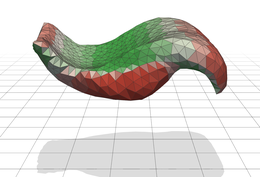}} & \raisebox{-.5\height}{\includegraphics[width=0.1575\linewidth]{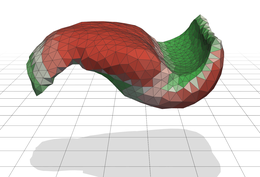}} & \raisebox{-.5\height}{\includegraphics[width=0.1575\linewidth]{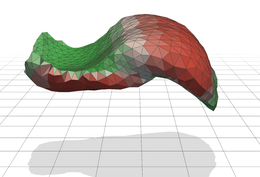}} \\
 & $\Delta x{=}{+}0.00$ & $\Delta x{=}{+}0.85$ & $\Delta x{=}{+}2.08$ & $\Delta x{=}{+}3.32$ & $\Delta x{=}{+}4.57$ & $\Delta x{=}{+}5.81$ \\
\end{tabular}}%

  \caption{Best 1k up--down swimmers of remesh-fTetWild and ours, rendered as in Figure~\ref{fig:frames}. Rows 1--2: design-stage wave, coloured by muscle group; rows 3--4: learned controller with $S{=}16$, coloured by muscle activation; $\Delta x$: forward displacement at time $t$; green dot: start position.}
  \label{fig:app-swim}
\end{figure}
Figure~\ref{fig:app-swim} plays back the best 1k up--down swimmers of remesh-fTetWild and ours under the design-stage wave, with amplitude 0.18, and under the learned controller with $S{=}16$, which drives 32 muscle blocks at its training amplitude of 0.24; Section~\ref{sec:exp:robot} gives their distances. The camera follows each body at a fixed offset, and the floor grid shows the distance travelled.

\section{Voxel and smooth surfaces}
\label{app:voxel}
\begin{table}[h]
  \centering\small\setlength{\tabcolsep}{3pt}
  \caption{Voxel re-evaluation of eight of our swimming designs from an earlier design sweep. Fitness: ours, the stored training fitness; voxel, median over three start phases; mean$\pm$sd over the $n$ designs; $\Delta$: change from ours; grid nodes relative to ours in brackets.}
  \label{tab:app-voxel}
  \begin{adjustbox}{max width=\textwidth}
\begin{tabular}{@{}llc cr crr crr@{}}
\toprule
 & & & \multicolumn{2}{c}{Ours (smooth)} & \multicolumn{3}{c}{Voxel, matched triangles} & \multicolumn{3}{c}{Voxel, $128^3$ grid} \\
\cmidrule(lr){4-5}\cmidrule(lr){6-8}\cmidrule(l){9-11}
Environment & Scale & $n$ & Fitness & Nodes & Fitness & $\Delta$ & Nodes & Fitness & $\Delta$ & Nodes \\
\midrule
swim, up--down & 1k & 3 & \best{$1.856 \pm 1.249$} & 1,965 & $0.830 \pm 0.617$ & $-55\%$ & 1,485 & $1.257 \pm 0.832$ & $-32\%$ & 28,420 (14.5$\times$) \\
swim, lateral & 1k & 3 & \best{$2.402 \pm 0.054$} & 2,026 & $1.261 \pm 0.040$ & $-48\%$ & 1,536 & $1.586 \pm 0.041$ & $-34\%$ & 29,913 (14.8$\times$) \\
swim, up--down & 8k & 1 & \best{$2.436$} & 11,712 & $1.535$ & $-37\%$ & 9,198 & $1.286$ & $-47\%$ & 30,818 (2.6$\times$) \\
swim, lateral & 8k & 1 & \best{$2.618$} & 12,008 & $1.327$ & $-49\%$ & 8,842 & $0.968$ & $-63\%$ & 32,548 (2.7$\times$) \\
\bottomrule
\end{tabular}
\end{adjustbox}
\end{table}
Table~\ref{tab:app-voxel} gives the voxel comparison of Section~\ref{sec:exp:robot} in full. It uses eight of our designs from an earlier sweep, distinct from the seeds of Table~\ref{tab:wave}: three per swimming condition at 1k and one at 8k. The smooth reference is the stored training fitness of each design, and each voxel body is scored by the median fitness over three start phases. Each design keeps its shape, muscle rule and fitness measure, and its tetrahedral body is replaced by a voxel body at one of two resolutions. The first has as many surface triangles as our mesh and 0.74--0.79$\times$ our node count; the second is the $128^3$ level-set grid of the design loop. The voxel bodies lose 32--63\% of the smooth designs' fitness.

\FloatBarrier
\section{Shape design with linear finite elements}
\label{app:fea}
\begin{figure}[t]
  \centering
{\small\setlength{\tabcolsep}{0pt}%
\begin{tabular}{@{}c@{\hspace{0.01\linewidth}}c@{\hspace{0.015\linewidth}}c@{}}
before (uniform section) & after (same volume) & \\
\raisebox{-.5\height}{\includegraphics[width=0.375\linewidth]{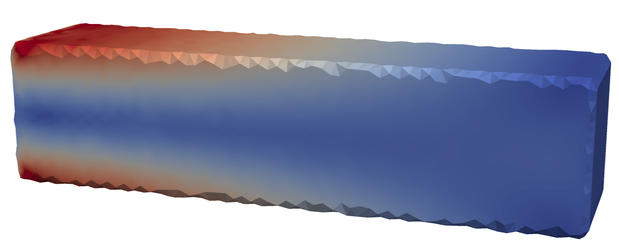}} & \raisebox{-.5\height}{\includegraphics[width=0.375\linewidth]{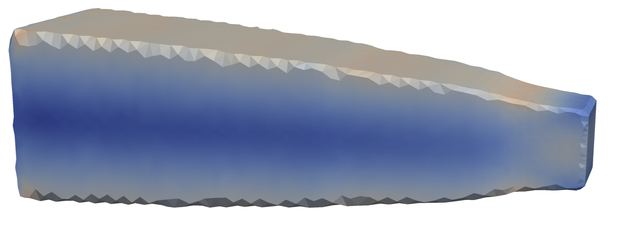}} & \raisebox{-.5\height}{\includegraphics[width=0.22\linewidth]{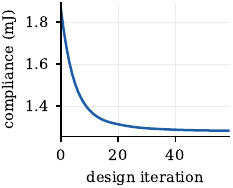}} \\
\end{tabular}}\\[4pt]
\includegraphics[width=\linewidth]{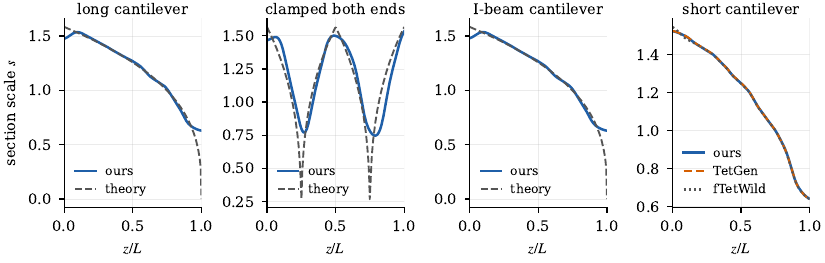}

  \caption{Top: compliance design of a short cantilever (1:1:4, tip load), uniform section and optimized shape at equal volume, coloured by von Mises stress, 0\,\raisebox{-0.15ex}{\includegraphics[width=3em,height=1.1ex]{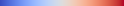}}\,5.3\,MPa; right: compliance per design step. Bottom: optimized section scale along the beam against the analytical profile; short cantilever: the same design on TetGen and fTetWild meshes.}
  \label{fig:app-fea}
\end{figure}
The meshes also support gradient-based design with linear finite elements. We minimize the compliance $C = \mathbf{f}^\top\mathbf{u}$ of a beam under a static 100\,N load at fixed volume, with aluminium properties ($E = 68.9$\,GPa, $\nu = 0.33$). The design variable is a section scale $s(z) \in [0.55, 1.75]$ along the beam axis, set by 12 segments through a sigmoid and blended by Gaussian weights along $z$; the cross-section is rescaled after each step so that the volume stays fixed. The cantilevers are clamped at the base and carry the load on the tip nodes; the beam clamped at both ends has both end faces pinned transversally and carries the load at mid-span. The gradient comes from the adjoint of the static solve, $\partial C / \partial p = -\mathbf{u}^\top (\partial K / \partial p)\,\mathbf{u}$, and Adam takes 60 steps. Our beam meshes use 8k interior spheres (11.5k--16.7k nodes), and the TetGen and fTetWild meshes of the short beams have 10.7k--12.9k nodes. Compliance falls by 30.4--34.1\% (Table~\ref{tab:app-fea}); on the short cantilever it drops from 1.86 to 1.28\,mJ, with 90\% of the drop reached by step 14. All four cantilevers taper from a section scale of 1.48--1.54 at the root to 0.63--0.64 at the tip, and their profiles have an RMS deviation of 0.067--0.069 from the uniform-strength profile $s \propto (L-z)^{1/3}$, at both aspect ratios. On the beam clamped at both ends the RMS deviation from $|M(z)|^{1/3}$ is 0.131, because the lower bound $s \geq 0.55$ holds the profile above zero where the bending moment $M$ vanishes (Figure~\ref{fig:app-fea}). On the two short beams, TetGen and fTetWild meshes of the same shape reach the same compliance reduction within 0.2 percentage points, and their profiles differ from ours by an RMS of 0.001--0.010.

\begin{table}[t]
  \centering\small\setlength{\tabcolsep}{5pt}
  \caption{Compliance before ($C_0$) and after ($C_1$) 60 design steps at fixed volume. Mesh: the volume mesher, at 10.7k--16.7k nodes.}
  \label{tab:app-fea}
\begin{tabular}{@{}l l r r r@{}}
\toprule
Structure & Mesh & $C_0$ (mJ) & $C_1$ (mJ) & $\Delta C$ \\
\midrule
Cantilever, 1:1:20 & ours & 210.86 & 139.10 & $-34.0\%$ \\
Cantilever, 1:1:4 & ours & 1.86 & 1.28 & $-31.1\%$ \\
 & TetGen & 1.85 & 1.27 & $-31.1\%$ \\
 & fTetWild & 1.83 & 1.26 & $-31.3\%$ \\
Clamped both ends, 1:1:20 & ours & 3.68 & 2.56 & $-30.4\%$ \\
I-beam cantilever, 1:1:20 & ours & 330.39 & 217.64 & $-34.1\%$ \\
I-beam cantilever, 1:1:4 & ours & 3.11 & 2.14 & $-31.2\%$ \\
 & TetGen & 3.08 & 2.12 & $-31.2\%$ \\
 & fTetWild & 2.96 & 2.04 & $-31.0\%$ \\
\bottomrule
\end{tabular}

\end{table}

\FloatBarrier
\FloatBarrier

\end{document}